\documentclass[letterpaper, 10 pt, conference]{ieeeconf}                                                       

\IEEEoverridecommandlockouts                              % This command is only
\usepackage{cite}
\usepackage[font=footnotesize]{caption}

\usepackage{subcaption}

\usepackage{amsmath,amssymb,amsfonts,bm}
\usepackage[bb=boondox]{mathalfa}
\usepackage{algorithmic}
\usepackage{graphicx}
\usepackage{textcomp}
\usepackage{xcolor}

\usepackage{tikz,tikz-3dplot}
\usetikzlibrary{3d,calc,shapes}

\usepackage{amsthm}

\newtheorem{proposition}{Proposition}

\usepackage{comment}
\usepackage{stfloats}
\usepackage{multirow}

\usepackage[
placement=top,
angle=0,
color=red,
scale=1.1,
hshift=0,
vshift=-15
]{background}
\backgroundsetup{contents=\bf{----------------------------------------------------------------- PREPRINT -----------------------------------------------------------------}}
\title{\LARGE \bf A Natural Indirect Adaptive Controller \\for a Satellite-Mounted Manipulator}
\author{Jacopo Giordano, Angelo Cenedese and Andrea Serrani% <-this % stops a space
\thanks{Jacopo Giordano and Angelo Cenedese are with the department of Information Engineering, University of Padua, Italy.
        Email: {\tt\footnotesize jacopo.giordano@phd.unipd.it},  {\tt\footnotesize cenedese@dei.unipd.it}
        }%
\thanks{Andrea Serrani is with the Department of  Electrical \& Computer Engineering, The Ohio State University, Columbus, OH 43210.
        Email: {\tt\footnotesize serrani.1@osu.edu}
        }%
\thanks{Corresponding Author: Jacopo Giordano.} 
}

\begin{document}

\maketitle
%\thispagestyle{empty}
%\pagestyle{empty}

%%%%%%%%%%%%%%%%%%%%%%%%%%%%%%%%%%%%%%%%%%%%%%%%%%%%%%%%%%%%%%%%%%%%%%%%%%%%%%%%

\begin{abstract}
The work considers the design of an indirect adaptive controller for a satellite equipped with a robotic arm manipulating an object. Uncertainty on the manipulated object can considerably impact the overall behavior of the system. In addition, the dynamics of the actuators of the base satellite are non-linear and can be affected by malfunctioning. Neglecting these two phenomena may lead to excessive control effort or degrade performance. An indirect adaptive control approach is pursued, which allows consideration of relevant features of the actuators dynamics, such as loss of effectiveness. 
Furthermore, an adaptive law that preserves the physical consistency of the inertial parameters of the various rigid bodies comprising the system is employed. 
The performance and robustness of the controller are first analyzed and then validated in simulation.  
\end{abstract}

\section{Introduction}

Space operations are continuously evolving and expanding in complexity and reach. In this very dynamic context, satellites equipped with a robotic arm, known as Space Manipulator Systems (SMSs), are expected to play a primary role in a large variety of operations, which can be roughly categorized into three groups: on-orbit servicing~\cite{OOS}, active debris removal~\cite{ADR} and on-orbit assembly and manufacturing~\cite{OOA}. 
A space manipulator system exhibits a strong dynamical coupling between the Base Satellite (BS) and the robotic arm, thereby requiring specific control solutions to achieve the required level of performance~\cite{papadopoulos2021robotic}. 
With the exclusion of the first type of mission, SMS operating in the aforementioned scenarios are subject to significant model uncertainty that is primarily linked to limited knowledge about the inertial parameters of the target. A further source of uncertainty stems from un-modeled nonlinearities in the actuator dynamics, which can be exacerbated by input saturation and component degradation, and may lead to mission failure if not taken into account in the control design.
%
%To control a system in the presence of disturbances and uncertainties two approaches are particularly popular: robust control and adaptive control. In the former, the presence of unmodelled behaviors is taken into account during the control design while considering a worst-case scenario. In the latter, the parameters of the controller or of the reference model are updated online during the control task in order to improve the control performance \cite{slotine1991applied}. In general robust control is usually preferred w.r.t. adaptive control if the knowledge about the nature of the uncertainties is limited and can not be captured by the control model. On the other hand, if uncertainties can be reasonably characterized by the control model, adaptive control is to be preferred since it usually achieves better performances with a lower control effort. Indeed, the ability to withstand to unmodelled behaviors of robust controllers usually comes at the price of a higher control effort and oscillations around the desired trajectory. 
%

In SMS scenarios, uncertainty related to the inertial parameters of the target can be adequately captured in a control-oriented model amenable to adaptive control strategies, due to regression models that are linear in the uncertain parameters. In principle, the effects of actuator saturation and wear can also be taken into account, for example by implementing anti-windup-like mechanisms and adding suitable estimates for losses of actuator efficiency. These additional features are better suited for indirect adaptive control techniques, whose application to SMS has not been yet reported to show difficulties in ensuring stable inversion of the estimated inertia matrix.  As a matter of fact, the totality of adaptive solutions for SMS reported in the literature considers a direct approach. One of the first adaptive controllers for a free-floating SMS adopts a direct computed-torque scheme using a dynamically equivalent manipulator model~\cite{parlaktuna2004adaptive}. That approach, however, requires acceleration feedback, which is difficult to obtain.  This issue is circumvented in the passivity-based adaptive controllers of~\cite{abiko2009adaptive,wang2009passivity}. In~\cite{wang2012prediction}  a modified task-space computed-torque controller is designed to counteract kinematic and dynamic uncertainties, and estimates are employed in lieu of the actual acceleration vector. Similarly,  in the task-space controller proposed in~\cite{hu2012adaptive},  the presence of external disturbances is explicitly considered within an adaptive backstepping control scheme that avoids the use of acceleration feedback. 
%
%The passivity-based adaptive controller of~\cite{yu2015modeling} employs a sliding mode observer to reconstruct the velocity vector. 
%
A different approach is proposed in \cite{christidi2020concurrent}, where parameter identification method based on angular momentum principles is coupled with a controller. Neural network-based controllers are considered in~\cite{yao2021adaptive}, where actuator saturations are explicitly accounted for. Nonetheless, the effectiveness of the proposed method depends on restrictive conditions. 

Notwithstanding a substantial research effort, the problem of dealing with significant model uncertainty and reduced actuator effectiveness remains an open problem in SMS applications. One reason is the lack of appropriate extensions of indirect adaptive control schemes, due to the necessity of guaranteeing physical consistency of the estimated inertial parameters for the manipulator and the target. To this end, this paper presents an indirect adaptive control scheme for a free-flying SMS aimed at solving a trajectory tracking problem in the joint space. Inspired by~\cite{lee2018natural}, a {\em natural adaptation law}, which preserves the underlying Riemannian geometry of inertial parameters, is employed for estimating the inertial parameters of the rigid bodies in the system. An observer-based dynamic estimator allows seamless incorporation of both input saturations and loss of control effectiveness of the actuators within the adaptive scheme. 
%
%
%\begin{itemize}
% \item A novel indirect Model Reference Adaptive Controller (MRAC). More in detail, the parameters of the reference model are updated via an adaptive observer, and the controller is designed starting from the adapted reference model. Theoretical guarantees about: the observation/tracking error convergence, the boundedness of the system trajectories, and the effects of external disturbances are provided.
% \item A natural adaptation law inspired from \cite{lee2018natural}, which respects the underlying Riemannian geometry of inertial parameters, is proposed for estimating the inertial parameters of the rigid bodies in the system. To the best of the authors' knowledge, none of the adaptive laws employed up to now in this context implicitly considers physical consistency in the parameters update. 
% \item An anti-windup-like mechanism is embedded into the adaptive observer to deal with input saturation and an efficiency adaptive term is added to take into account potential performance loss of the actuators.
%\end{itemize}

%struttura paper
The paper is organized as follows: Section~\ref{sec:AssAndNot} presents the main assumptions and sets the notation. In Section~\ref{sec:SysMod}, the dynamic model of the SMS and its actuators are presented, together with background notions on the space of physically consistent inertial parameters of a rigid body. The design of the adaptive controller is presented in Section~\ref{sec:AdaptContr}, together with a stability analysis. Simulation results are presented and discussed in Section~\ref{sec:Sim}, whereas concluding remarks are offered in Section~\ref{sec:Conc}.

%%%%%%%%%%%%%%%%%%%%%%%%%%%%%%%%%%%%%%%%%%%%%%%%%%%%%%%%%%%%%%%%%%%%%%%%%%%%%%%%
\section{Assumptions and Notation}
\label{sec:AssAndNot}
\subsection{System and Environmental Assumptions}
A SMS equipped with a single robotic arm comprising all revolute joints is considered in this work. 
%The manipulator joints are supposed to be all revolute since prismatic ones are not commonly employed in space. 
Flexibility of bodies and joints is assumed to be negligible. It is also assumed that the manipulation tasks are performed at a faster time scale than the orbital period, so that the influence of orbital dynamics and gravitational disturbances is neglected. The SMS is thus modeled as a chain of $n+1$ rigid bodies, arranged from the BS to the end effector (EE). The uncertain inertial parameters of the rigid bodies are assumed to range within known compact sets. Without loss of generality, the grasped object is assumed to be part of the EE. 
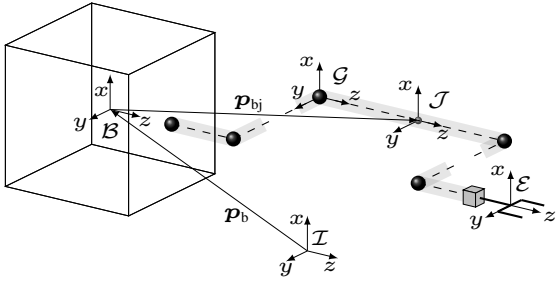
\begin{figure}[t!]
\centering
    \tdplotsetmaincoords{70}{125}

\begin{tikzpicture}[
		tdplot_main_coords,
		axis/.style={->,black,very thin,-latex},
		cube/.style={very thin},
        link/.style={line width=2mm,opacity=0.1},
        every node/.style={inner sep=1pt,scale=1.6}]
    
	%draw the axes

	\draw[axis] (-5,0,-3) -- (-5,0.5,-3) node[yshift=-2pt,xshift=-2pt]{\tiny $z$};
	\draw[axis] (-5,0,-3) -- (-4.5,0,-3) node[yshift=-2pt,xshift=0pt]{\tiny $y$};
	\draw[axis] (-5,0,-3) -- (-5,0,-2.5) node[yshift=-1pt,xshift=-2.5pt]{\tiny $x$};
    \draw (-5,0,-3) node[yshift=3pt,xshift=3pt]{\tiny $\mathcal{I}$};
 
	\draw[axis] (1,1,1) -- (1.5,1,1) node[yshift=-1pt,xshift=-2pt]{\tiny $y$};
	\draw[axis] (1,1,1) -- (1,1.5,1) node[yshift=-1pt,xshift=1pt]{\tiny $z$};
	\draw[axis] (1,1,1) -- (1,1,1.5) node[yshift=-4pt,xshift=-2pt]{\tiny $x$};
    \draw (1,1,1) node[yshift=-5pt,xshift=0pt]{\tiny $\mathcal{B}$};

    \draw[axis] (-1,3,1) -- (-1,3.5,1) node[yshift=2pt,xshift=1pt]{\tiny $z$};
	\draw[axis] (-1,3,1) -- (-1,3,1.5) node[yshift=0pt,xshift=-3pt]{\tiny $x$};
    \draw[axis] (-1,3,1) -- (-0.4,3,1) node[yshift=4pt,xshift=1pt]{\tiny $y$};
    \draw (-1,3,1) node[yshift=5pt,xshift=5pt]{\tiny $\mathcal{G}$};%_{\hspace{-1pt} j}$};

    \draw[axis] (-1,4.6,1) -- (-1,5,1) node[yshift=-2pt,xshift=0pt]{\tiny $z$};
	\draw[axis] (-1,4.6,1) -- (-1,4.6,1.5) node[yshift=0pt,xshift=-3pt]{\tiny $x$};
    \draw[axis] (-1,4.6,1) -- (-0.4,4.6,1) node[yshift=-2pt,xshift=0pt]{\tiny $y$};
    \draw (-1,4.6,1) node[yshift=5pt,xshift=5pt]{\tiny $\mathcal{J}$};%_{\hspace{-1pt} j}$};
    \shade[ball color=black!50] (-1,4.6,1) circle (.05cm);

    \draw[axis] (1,7.5,1) -- (1,8,1) node[yshift=-1pt,xshift=2pt]{\tiny $z$};
	\draw[axis] (1,7.5,1) -- (1.6,7.5,1) node[yshift=-1pt,xshift=-2pt]{\tiny $y$};
    \draw[axis] (1,7.5,1) -- (1,7.5,1.5) node[yshift=0pt,xshift=-3pt]{\tiny $x$};
    \draw (1,7.5,1) node[yshift=5pt,xshift=3pt]{\tiny $\mathcal{E}$};

    % Cube
	%draw the front-right of the cube
	\draw[cube] (0,0,0) -- (0,2,0) -- (2,2,0) -- (2,0,0) -- cycle;	
	%draw the back-right of the cube
	\draw[cube] (0,0,0) -- (0,2,0) -- (0,2,2) -- (0,0,2) -- cycle;
	%draw the back-left of the cube
	\draw[cube] (0,0,0) -- (2,0,0) -- (2,0,2) -- (0,0,2) -- cycle;
	%draw the front-right of the cube
	\draw[cube] (2,0,0) -- (2,2,0) -- (2,2,2) -- (2,0,2) -- cycle;
	%draw the front-left of the cube
	\draw[cube] (0,2,0) -- (2,2,0) -- (2,2,2) -- (0,2,2) -- cycle;
	%draw the top of the cube
	\draw[cube] (0,0,2) -- (0,2,2) -- (2,2,2) -- (2,0,2) -- cycle;

    % Links
    \draw[link] (1,2,1) -- (1,3,1);
    \draw[dashed] (1,2,1) -- (1,3,1);
    
    \draw[link] (1,3,1) -- (0.3,3,1);
    \draw[dashed] (1,3,1) -- (0.1,3,1);

    \draw[dashed] (-0.2,3,1) -- (-1,3,1);
    \draw[link] (-0.5,3,1) -- (-1,3,1);
 
    \draw[link] (-1,3,1) -- (-1,6,1);
    \draw[dashed] (-1,3,1) -- (-1,6,1);
    
    \draw[link] (1,6,1) -- (0.3,6,1);
    \draw[dashed] (1,6,1) -- (0.1,6,1);

    \draw[dashed] (-0.3,6,1) -- (-1,6,1);
    \draw[link] (-0.5,6,1) -- (-1,6,1);
    
    \draw[link] (1,6,1) -- (1,7,1);
    \draw[dashed] (1,6,1) -- (1,7,1);

    % Joints
    \shade[ball color=black] (1,2,1) circle (.1cm);
    \shade[ball color=black] (1,3,1) circle (.1cm);
    \shade[ball color=black] (-1,3,1) circle (.1cm);
    \shade[ball color=black] (-1,6,1) circle (.1cm);
    \shade[ball color=black] (1,6,1) circle (.1cm);
    %\shade[ball color=black] (1,7,1) circle (.1cm);

    % End Effector
    \draw[cube,black,fill=gray!50] (0.9,7,0.9) -- (1.1,7,0.9) -- (1.1,7,1.1) -- (0.9,7,1.1) -- cycle;	
    \draw[cube,black,fill=gray!50] (0.9,7,1.1) -- (0.9,6.8,1.1) -- (1.1,6.8,1.1) -- (1.1,7,1.1) -- cycle;  
    \draw[cube,black,fill=gray!50] (1.1,6.8,1.1) -- (1.1,6.8,0.9) -- (1.1,7,0.9) -- (1.1,7,1.1) -- cycle; 
    
    \draw[thick] (1,7,1) -- (1,7.5,1);
    \draw[thick] (1.3,7.5,1) -- (0.75,7.5,1);
    \draw[thick] (1.3,7.5,1) -- (1.3,7.9,1);
    \draw[thick] (0.75,7.5,1) -- (0.75,7.87,1);

    %arrow
    \draw[axis] (-5,0,-3) -- (1,1,1) node[yshift=-25pt,xshift=30pt]{\tiny $\bm{p}_{\scalebox{.7}{b}}$};
    %\draw[axis] (1,1,1) -- (-1,2.95,1);
    \draw[axis] (1,1,1) -- (-1,4.6,1) node[yshift=4pt,xshift=-40pt]{\tiny $\bm{p}_{\scalebox{.7}{bj}}$};

\end{tikzpicture}
    \caption{Structure of the SMS with the indication of the frames of interest.  }
    \label{fig:SM}
\end{figure}
\subsection{Notation} 
The reference frames employed throughout this work are shown in Fig.\ref{fig:SM}, where $\mathcal{I}$ denotes the inertial frame. For the $j^{th}$ rigid body of the kinematic chain, with $j=0,\dots,n$, a generic frame $\mathcal{G}$ is positioned in a geometrically significant pose, and another frame $\mathcal{J}$ is placed at its center-of-mass (CoM) and aligned with its principal axes of inertia. The frame $\mathcal{G}$ of the BS is placed at its geometric center. 
% and is aligned with the solar panels and the launch adaptor. 
For each link of the robotic arm, the frame $\mathcal{G}$ is positioned at the joint center and aligned accordingly to the Denavit-Hartenberg convention. As the inertial parameters are not exactly known, the frames $\mathcal{J}$ cannot be defined a priori, but must be specified from the frames $\mathcal{G}$ using estimates of the inertial parameters. A body frame $\mathcal{B}$ aligned with the $j=0$ frame $\mathcal{G}$, and a frame $\mathcal{E}$ specifying the EE pose are defined for convenience.       
Given three generic frames $\mathcal{M}$, $\mathcal{N}$ and $\mathcal{O}$, the vector $\bm{p}^o_{mn} \in \mathbb{R}^3$ represents the position of the origin of frame $\mathcal{N}$ relative to the origin of $\mathcal{M}$ expressed in $\mathcal{O}$. Similarly, the linear and angular velocities of $\mathcal{N}$ relative to $\mathcal{M}$ resolved in $\mathcal{O}$  are denoted  with $\bm{v}^o_{mn} \in \mathbb{R}^3$ and $\bm{w}^o_{mn} \in \mathbb{R}^3$, respectively. The rotation matrix needed to align $\mathcal{N}$ with $\mathcal{M}$ is represented with the symbol $\bm{R}_{mn} \in SO(3)$. %The same subscript notation is used for unit quaternions $\bm{q}_{mn} = \left[\eta_{mn} \ \bm{\epsilon}^T_{mn} \right]^T \in \mathbb{S}^3$ where $\eta$ and $\bm{\epsilon}$ are its scalar and vectorial part respectively. The quaternion Hamiltonian product is symbolized with $\otimes$. 
The vectors $\bm{f}^o_{m} \in \mathbb{R}^3$ and $\bm{\tau}^o_{m} \in \mathbb{R}^3$ represent forces and torques applied at the origin of frame $\mathcal{M}$ and expressed in $\mathcal{O}$. The subscript $\cdot_{ref}$ is added to variables that are related to reference trajectories. 
The mass and inertia matrix of the $j^{th}$ rigid body expressed in frame $\mathcal{O}$ are denoted by the symbols $m_{j}$ and $\bm{I}_{j}^o \in \mathbb{R}^{3 \times 3}$, respectively. Body frame is used whenever the superscript representing the reference frame is not specified. Conversely, if a subscript is not specified, then the first frame is replaced by the inertial one. Finally, the $ n\times n$ identity matrix and zero matrix are respectively denoted by $\mathbb{I}_{n \times n}$ and $\mathbb{0}_{n \times n}$.

\section{System Dynamics} 
\label{sec:SysMod}
The dynamical model of the system expressed in body-frame  coordinates is given as follows (see~\cite{giordano2019coordinated} for details)
%in such coordinates the equations do not depend neither on the position or on the orientation of the BS. Using a compact matrix form, the equations of the dynamical model can be written as follows:  

\begin{equation}
    \begin{split}
    \underbrace{ 
    \begin{bmatrix}
    \bm{M}_t & \bm{M}_{tr} & \bm{M}_{tm}\\
    \bm{M}_{tr}^T & \bm{M}_r & \bm{M}_{rm}\\
    \bm{M}_{tm}^T & \bm{M}_{rm}^T & \bm{M}_m\\
    \end{bmatrix}
    }_{\bm{M}(\bm{q})}
    \begin{bmatrix}
    \dot{\bm{v}}_{b}\\
    \dot{\bm{w}}_{b}\\
    \ddot{\bm{q}}
    \end{bmatrix}
    + \\
    \underbrace{ 
    \begin{bmatrix}
    \bm{C}_t & \bm{C}_{tr} & \bm{C}_{tm}\\
    \bm{C}_{tr}^T & \bm{C}_r & \bm{C}_{rm}\\
    \bm{C}_{tm}^T & \bm{C}_{rm}^T & \bm{C}_m\\
    \end{bmatrix}
    }_{\bm{C}(\bm{q},\bm{v}_{b},\bm{w}_{b},\dot{\bm{q}})}
    \begin{bmatrix}
    \bm{v}_{b}\\
    \bm{w}_{b}\\
    \dot{\bm{q}}
    \end{bmatrix} 
    =
    \begin{bmatrix}
    \bm{f}_{b}\\
    \bm{\tau}_{b}\\
    \bm{\tau}
    \end{bmatrix} 
    \end{split}
    \label{eq:DynMod}
\end{equation}
where $\bm{q}\in \mathbb{R}^n$ are the positions of the arm joints, $\bm{\tau}\in \mathbb{R}^n$ represents the torques applied at the arm joints, $\bm{M}(\bm{q}) \in \mathbb{R}^{(6+n)\times (6+n)}$ is the inertia matrix and $\bm{C}(\bm{q},\bm{v}_{b},\bm{w}_{b},\dot{\bm{q}}) \in \mathbb{R}^{(6+n)\times (6+n)}$ is the centrifugal/Coriolis matrix. 
%In greater detail, the single blocks that make up $\bm{M}$ are:
%\begin{subequations}
%    \begin{align}
%   &\bm{M}_t = \sum_{j=0}^n m_j \mathbb{I}_{3 \times 3}  \in \mathbb{R}^{3 \times 3} \\
%   &\bm{M}_{tr} = \sum_{j=1}^n [\bm{p}_{bj}]_\times m_j \\
%   &\bm{M}_{tm} = \sum_{j=1}^n m_i \bm{J}_{\bm{v}_j} \in \mathbb{R}^{3 \times n}\\
%   &\bm{M}_r = \bm{I}_b + \sum_{j=1}^n \left( \bm{I}_j - m_j [\bm{p}_{bj}]_\times [\bm{p}_{bj}]_\times \right) \in \mathbb{R}^{3 \times 3} \\ 
%   & \bm{M}_{rm} = \sum_{j=1}^n \left( m_j \bm{J}_{{\bm{\omega}}_j} + m_j [\bm{p}_{bj}]_\times \bm{J}_{\bm{v}_j} \right) \in \mathbb{R}^{3 \times n}\\
%   & \bm{M}_{m} = \sum_{j=1}^n \left( \bm{J}_{\bm{\omega}_j}^T \bm{I}_j \bm{J}_{\bm{\omega}_j} + m_j \bm{J}_{\bm{v}_j}^T \bm{J}_{\bm{v}_j} \right) \in \mathbb{R}^{n \times n}
%   \end{align}
%\end{subequations}
%
%where $\bm{J}_{\bm{v}_j}(\bm{q}) \in \mathbb{R}^{3 \times n}$ and $\bm{J}_{\bm{\omega}_j}(\bm{q}) \in \mathbb{R}^{3 \times n}$ are the linear and angular velocity Jacobians of the manipulator respectively and $[\cdot]_\times \in \mathbb{R}^{3  \times 3 }$ is the skew symmetric matrix operator. It is not worth reporting the explicit formulations of each of the Coriolis $\bm{C}$ matrix blocks due to the complexity of their expressions. 
%
To simplify notation,  define the generalized velocities and forces as $\dot{\bm{x}}^T= \left[ \bm{v}_{b}, \bm{w}_{b}, \dot{\bm{q}} \right]^T \in \mathbb{R}^{6+n} $ and $\bm{u}^T= \left[ \bm{f}_{b}, \bm{\tau}_{b}, \bm{\tau} \right]^T \in \mathbb{R}^{6+n}$.

\subsection{Actuators Dynamics}
The actuation devices present on an SMS can be divided into two groups: one related to the robotic arm, and the other to the BS.  The joints of the robotic arm are actuated by electric motors equipped with gearboxes. The BS is usually equipped with a Reaction Control System (RCS) and Reaction Wheels (RWs). A RCS typically consists of a cluster of one-sided jet thrusters, operating in off-on mode. 
%
%There exist proportional thrusters that are able to modulate the amount of force produced by proportionally opening the fuel valve. However, they are not commonly used due to their complexity and the tendency of the fuel valve to get jammed. Consequently, the majority of jet thrusters are used in on-off mode in space applications. 
%
A continuous thrust command must be modulated by a train of discrete pulses that on average produce the desired outcome (e.g., using Pulse Width Modulation (PWM), Pulse-Width Pulse-Frequency (PWPF)). The minimum amount of force that can be generated depends on the minimum opening time of the fuel valve. In addition, force and torque commands need to be mapped onto the set of thrusters. 
A RW consists of an electric motor controlling a flywheel, producing torque along the axis of rotation.
%
%Hence, its dynamics correspond to the dynamics of the the electric motor employed to spin the flywheel.
%Following a similar reasoning, the dynamics of the manipulator joints depend on the one of the electric motor and gearbox used. 
%
Consequently, 
\begin{equation}
    \bm{f}_b = \bm{f}_b^{th}, \quad \bm{\tau}_b = \bm{\tau}_b^{th} + \bm{\tau}_b^{rw}, \quad \bm{\tau} = \bm{\tau}^{em+gear} 
\end{equation}
where the terms on the right-hand side embed all the above-mentioned dynamics. The superscripts $\cdot^{th}$, $\cdot^{rw}$, $\cdot^{em+gear}$ denote thrusters, RW, and motors and gearboxes of the joints. 
%
%
%Furthermore, it is possible to capture a potential 
Performance degradation of the actuators is modeled by modifying the right-hand side of equation~\eqref{eq:DynMod} as follows:
\begin{equation}
    \bm{M}(\bm{q}) \ddot{\bm{x}} + \bm{C}(\bm{q},\dot{\bm{x}}) \dot{\bm{x}} = \mathrm{diag} (\bm{\lambda}_{act} ) \bm{u} 
    \label{eq:DynModEff}
\end{equation}
where $\mathrm{diag}(\cdot)$ maps a vector into a diagonal matrix, and $\bm{\lambda}_{act} = \left[ \lambda_{act,1}, \dots, \lambda_{act,6+n} \right]^T$  is an efficiency vector in the convex compact set $ \Lambda_{act} := \{\bm{\lambda} \in \mathbb{R}^{6+n} \vert \lambda_{min} \leq \lambda_{i} \leq 1 \ \mathrm{with} \ i = 1,\dots ,6+n \ \mathrm{and} \ \lambda_{min} \in (0,1] \}$.    

\subsection{Linear Formulation w.r.t. Inertial Parameters}
%Similarly to what can by done in the case of a ground manipulator, it is possible to rewrite the equations of the dynamical model \eqref{eq:DynModEff} in a form that depends linearly on the on the inertial parameters of its rigid bodies: 
It is well known that the model~\eqref{eq:DynModEff} admits a linear parameterization of the form
\begin{equation}
    \bm{Y}(\bm{q},\dot{\bm{x}},\ddot{\bm{x}}) \bm{\theta} = \mathrm{diag} (\bm{\lambda}_{act} ) \bm{u} 
    \label{eq:LinDynModEff}
\end{equation}
where $\bm{\theta}=\left[\bm{\theta}_0^T,\dots,\bm{\theta}_{n}^T \right]^T \in \Theta$ collects the inertia parameters of the $n+1$ rigid bodies, $\Theta = \Pi_{j=0}^{n} \Theta_j, \ \Theta_j  \subset \mathbb{R}^{10}$  is the product space of $n+1$ known compact convex sets, and $\bm{Y}(\bm{q},\dot{\bm{x}},\ddot{\bm{x}}) \in \mathbb{R}^{(6+n) \times 10(n+1)}$ is a regressor. %Such formulation is essential for designing an adaptive controller. 
%
%Such formulation is very useful and can be employed to tackle a great variety of problems. For example, it is essential in the design process of adaptive controllers and it can be also used to identify/estimate the inertial parameters of the system. 
%
The $j^{th}$ rigid body inertial parameters are
\begin{equation}
    \bm{\theta}_j = \left[ m,\bm{h}_c^T, I^{xx}, I^{yy}, I^{zz}, I^{xy}, I^{yz}, I^{zx} \right]^T \in \Theta_j 
    \label{eq:inertialParam}
\end{equation}
where $m := m_j \in \mathbb{R}$, $\bm{h}_c := m_j \bm{p}^j_{gj} \in \mathbb{R}^{3}$ is the product between $m$ and the distance from the $j^{th}$ joint to the $j^{th}$ link center of mass, and $I^{\alpha \alpha} \in \mathbb{R}$ are the elements of the inertia matrix $\bm{I}_j^j \in \mathbb{R}^{3 \times 3}$. For physically consistent rigid bodies, the vector $\bm{\theta}_j$ does not span the entirety of $\mathbb{R}^{10}$ but only a subset of it. As explained in \cite{wensing2017linear}, the various requirements necessary for physical consistency, can be condensed in single positivity constraint on the symmetric matrix $\bm{P}_i \in \mathcal{S}(4)$ obtained from $\bm{\theta}_i$ using the following bijective linear map:
\begin{equation}
    f(\bm{\theta}_i) = \bm{P}_i = \begin{bmatrix}
    0.5 \ \mathrm{tr}\left(\bm{I}_j^j\right)\ \mathbb{I}_{3 \times 3}- \bm{I}_j^j & \bm{h}_c\\
    \bm{h}_c^T & m
    \end{bmatrix} \in \mathcal{S}(4)
    \label{eq:mapf}
\end{equation}
\begin{equation}
    f^{-1}(\bm{P}_i) = \begin{bmatrix}m, \bm{h}_c^T , \mathrm{tr}(\bm{\Sigma}_i)\ \mathbb{I}_{3 \times 3} - \bm{\Sigma}_i \end{bmatrix}^T \in \mathbb{R}^{10}
    \label{eq:mapfinv}
\end{equation}
where $\bm{\Sigma}_j = 0.5 \mathrm{tr}\left(\bm{I}_j^j\right) \mathbb{I}_{3 \times 3} - \bm{I}_j^j $ and $\mathcal{S}(4)$ is set of the $ 4\times 4$ symmetric matrices. 
%
%Using what is stated above, it is possible to define 
The convex set $\mathcal{M}$ of physically consistent parameters for a rigid body is defined as~\cite{lee2018geometric}
\begin{equation}
    \begin{split}
        \mathcal{M} & = \{ \bm{\theta}_i \in \mathbb{R}^{10}:  f(\bm{\theta}_i) \succ 0 \} \subset \mathbb{R}^{10} \\
        & \simeq \{ \bm{P}_i \in \mathcal{S}(4):  \bm{P}_i \succ 0 \} = \mathcal{P}(4)
    \end{split}
\end{equation}
where $\mathcal{P}(4)$ is the set of positive definite symmetric matrices that is a submanifold of $\mathcal{S}(4)$. When considering the $n+1$ bodies system as a whole, the set of physically consistent parameters is given by the product space $\mathcal{M}^{n+1} \simeq \mathcal{P}^{n+1}(4)$ which is convex as well. 
In addition, it is worth defining for later use the set $\mathcal{M}_c = \mathcal{M} \cap \Theta_i$ that is the set of physically consistent inertial parameters inside the compact set $\Theta_i$. As before, when considering $n+1$ rigid bodies, the product space $\mathcal{M}_c^{n+1}$ will be used.

\subsection{Riemannian Geometry of $\mathcal{M} \simeq \mathcal{P}(4)$}
The manifold $\mathcal{M}$ inherits the Riemannian structure of $\mathcal{P}(4)$, due to the characteristics of the map $f(\cdot)$ and its inverse. As discussed in \cite{lee2018geometric}, this allows a coordinate-invariant distance on $\mathcal{M}$ to be defined starting from the affine invariant metric on $\mathcal{P}(4)$. Since $\mathcal{P}(4)$ is an open subset of $\mathcal{S}(4)$, it is possible to define for each $\bm{P} \in \mathcal{P}(4)$ the tangent space $T_{\bm{P}} \in T\mathcal{S}(4)$ at $\bm{P}$. Given $\bm{P} \in \mathcal{P}(4)$ and $\bm{X},\bm{Y} \in T_{\bm{P}}$ % and using the above mentioned affine invariant metric it is possible to define the following inner product:
the inner product
\begin{equation}
    \langle \bm{X},\bm{Y} \rangle_P = \frac{1}{2} \mathrm{ tr}(\bm{P}^{-1} \bm{X} \bm{P}^{-1} \bm{Y})
    \label{eq:innerProduct}
\end{equation}
%where $\mathrm{tr}(\cdot)$ is the trace operator. This inner product 
is invariant w.r.t. the group action $\bm{Q} \ast \bm{P} = \bm{Q} \bm{P} \bm{Q}^T$ with $\bm{Q} \in GL(4)$ namely the set of $4 \times 4$ invertible matrices.
It is shown in~\cite{lee2018geometric} that the Riemannian distance between points $\bm{P}_1$ and $\bm{P}_2$ in $\mathcal{P}(4)$ is
\begin{equation}\label{eq:distance}
%    \begin{split}
    d_{\mathcal{P}(4)}(\bm{P}_1, \bm{P}_2) =  \vert \vert \log \left( \bm{P}_1^{-1} \bm{P}_2 \right) \vert \vert_F   %= \left( \sum^4_{i=1} \left( \log (\lambda_i) \right)^2 \right)^{\frac{1}{2}}
%    \end{split}
\end{equation}
where $\lambda_i$ are the eigenvalues of $\bm{P}_1^{-1}\bm{P}_2$ or equivalently $\bm{P}_1^{-\frac{1}{2}}\bm{P}_2 \bm{P}_1^{-\frac{1}{2}}$. 
Finally, using the map $f(\cdot)$ in~\eqref{eq:mapf} and the distance %in $\mathcal{P}(4)$ defined 
in~\eqref{eq:distance}, a metric on $\mathcal{M}$ can be defined as
\begin{equation}\label{eq:metric}
    d_{\mathcal{M}}(\bm{\theta}_1,\bm{\theta}_2)=d_{\mathcal{P}(4)}(f(\bm{\theta}_1),f(\bm{\theta}_2)) 
\end{equation}
where $\bm{\theta}_1,\bm{\theta}_2 \in \mathcal{M}$. Noteworthy properties of the metric~\eqref{eq:metric}, among several discussed in \cite{lee2018geometric}, are invariance with respect to coordinate frames, and physical/scale invariance.  

\subsection{Approximation of the Distance Metric on $\mathcal{M}$}
The metric on $\mathcal{P}(4)$ defined in~\eqref{eq:distance} is not suitable to be used as a Lyapunov function in the context of adaptive control, due to its nonlinearity. In \cite{lee2018natural}, a better-suited pseudo-distance metric based on the Bregman divergence of the function  $F(\cdot) = -\log (\det (\cdot))$ on \textbf{$\mathcal{P}(4)$} is proposed. To this end, consider a continuously differentiable, strictly convex function $F: \Omega \rightarrow \mathbb{R}$ defined on a convex set $\Omega$. The Bregman divergence of $F$ of points $p,q \in \Omega$, namely
\begin{equation}
    D_{F(\Omega)}(p \vert \vert q) = F(p)-F(q)- \langle \nabla F(q), p-q \rangle
\end{equation}
can be interpreted as the difference between the value of $F(\cdot)$ at point $p$ and the value of the first order Taylor expansion of $F(\cdot)$ around $q$ evaluated at $p$. Therefore, if $\Omega = \mathcal{P}(4)$ and $F(\cdot) = -\log (\det (\cdot))$ then the Bregman divergence is:
\begin{align}\label{eq:bregman}
     D_{F(\mathcal{P}(4))}(\bm{P}_2 \vert \vert \bm{P}_1) & =  \log \left( \frac{\det(\bm{P}_1)}{\det(\bm{P}_2)} \right)  + \mathrm{tr}\left(\bm{P}_1^{-1} \bm{P}_2 \right) - 4 \nonumber \\
                                                  & = \sum^4_{i=1} \left( - \log (\lambda_i) + \lambda_i - 1 \right)
\end{align}
where $\bm{P}_1, \bm{P}_2 \in \mathcal{P}(4)$, and $\lambda_i$ are the eigenvalues of $\bm{P}_1^{-1}\bm{P}_2$, or equivalently $\bm{P}_1^{-\frac{1}{2}}\bm{P}_2 \bm{P}_1^{-\frac{1}{2}}$.
The Bregman divergence being a presudo-distance fails 
%to be symmetric and 
to satisfy the triangular inequality; however both positivity and coordinate invariance are preserved. 
%The two lost properties do not impede the use of this pseudo-distance in the design of an adaptive controller.
Similarly as before, a pseudo-distance metric on $\mathcal{M}$ can be defined using the map $f(\cdot)$ in~\eqref{eq:mapf} and the Bregman divergence
\begin{equation}
    D_{\mathcal{M}}(\bm{\theta}_1 \vert \vert \bm{\theta}_2) = D_{F(\mathcal{P}(4))}(f(\bm{\theta}_2) \vert \vert f(\bm{\theta}_1))
\end{equation}
It is possible to prove that  $D_{\mathcal{M}}(\bm{\theta}_1 \vert \vert \bm{\theta}_2)$ approximates the metric $d_{\mathcal{M}}(\bm{\theta}_1,\bm{\theta}_2)$ in~\eqref{eq:metric} up to second-order~\cite{lee2018natural}.

\section{Indirect Adaptive Controller} 
\label{sec:AdaptContr}
%As mentioned in the introduction, the rationale behind adaptive control is to dynamically update the parameters of the controller or of the reference model in order to make the system follow a certain trajectory while keeping the tracking error small. In general, there are many types of adaptive controllers. 
%
In this section, an indirect  adaptive controller is designed to achieve asymptotic trajectory tracking for the SMS, in spite of model parameter uncertainty and loss of actuator efficiency. In the proposed implementation, estimates of the  parameters of the model are updated using an adaptive observer, giving rise to a dynamic parameter estimator. The adaptive observer is then used to design the controller. The advantage of using an indirect approach for parameter estimation lies in the modularity of the scheme and advantageous robustness properties in presence of actuator dynamics, whenever these latter are adequately modeled and incorporated in the dynamic estimator. In the context of robotic manipulation, however, the use of an adaptive observer is hindered by the requirement to invert the estimated inertia matrix, which may fail to maintain positive definiteness as the estimated parameters evolve. A projection operator defined on the basis of the pseudo-Riemannian metric thus presented is employed in the update law to ensure consistency of the estimated inertia in the space of positive definite matrices. 
\subsection{Adaptive Observer}
%In a real-world scenario, it is not possible to know a priori the real values of $\bm{\theta}$ and $\bm{\lambda}_{act}$. Indeed, the purpose of the adaptive observer presented next is to continuously update their estimates, which are represented with $\hat{\bm{\theta}}$ and $\hat{\bm{\lambda}}_{act}$ respectively, with the purpose of making the velocity observation error $\dot{\tilde{\bm{x}}}_{obs} = \dot{\bm{x}}-\dot{\hat{\bm{x}}}_{obs}$ go to zero. To obtain this result, it is not necessary to bring their identification errors, which are represented with $\tilde{\bm{\theta}}=\bm{\theta}-\hat{\bm{\theta}}$ and $\tilde{\bm{\lambda}}_{act}=\bm{\lambda}_{act}-\hat{\bm{\lambda}}_{act}$ respectively, to zero but it is sufficient to find some values that accordingly to the excited dynamics lead $\dot{\tilde{\bm{x}}}_{obs}$ to zero.   
%
%It is convenient to define now the adapted reference model (ARM) and the reference model error (RME). 
To define the adaptive observer, we start by expressing the left-hand side of equation \eqref{eq:LinDynModEff} in terms of the inertial parameter estimates $\hat{\bm{\theta}}$ and the ensuing estimation error $\tilde{\bm{\theta}}=\bm{\theta}-\hat{\bm{\theta}}$. Since \eqref{eq:DynModEff} and \eqref{eq:LinDynModEff} are equivalent, it follows that
\begin{subequations}
\begin{align}
    \bm{Y}(\bm{q},\dot{\bm{x}},\ddot{\bm{x}}) \bm{\theta} & %= \bm{Y}(\bm{q},\dot{\bm{x}},\ddot{\bm{x}}) (\hat{\bm{\theta}} + \tilde{\bm{\theta}})\\ 
    = \bm{Y}(\bm{q},\dot{\bm{x}},\ddot{\bm{x}}) \hat{\bm{\theta}} + \bm{Y}(\bm{q},\dot{\bm{x}},\ddot{\bm{x}}) \tilde{\bm{\theta}} \label{eq:eq19a}\\ 
    & = \bm{M}(\bm{q},\hat{\bm{\theta}}) \ddot{\bm{x}} + \bm{C}(\bm{q},\dot{\bm{x}},\hat{\bm{\theta}}) \dot{\bm{x}} \label{eq:eq19b} \\
    & \ \quad + \bm{M}(\bm{q},\tilde{\bm{\theta}}) \ddot{\bm{x}} + \bm{C}(\bm{q},\dot{\bm{x}},\tilde{\bm{\theta}}) \dot{\bm{x}} \label{eq:eq19c}
\end{align}
\end{subequations}
It should be clear that the terms in \eqref{eq:eq19b} and \eqref{eq:eq19c} are linearly parameterized in $\hat{\bm{\theta}}$ and $\tilde{\bm{\theta}}$, respectively.
%the two terms in \eqref{eq:eq19a} are linearly dependent versions w.r.t. the inertial parameters of the ADR and the RME respectively. 
%The same two terms but in their classical non-linear versions are reported in \eqref{eq:eq19c} and \eqref{eq:eq19d} respectively where their dependency on $\hat{\bm{\theta}}$ and $\tilde{\bm{\theta}}$ has been highlighted. 
Furthermore, $\bm{M}(\bm{q},\hat{\bm{\theta}})$ is positive definite (hence invertible) if $\hat{\bm{\theta}}$ is physically consistent. The adaptive observer used to generate a regressor for $\hat{\bm{\theta}}$ and the estimate $\hat{\bm{\lambda}}_{act}$ of $\bm{\lambda}_{act}$ is designed as
\begin{equation}
\begin{split}
    \bm{M}(\bm{q},\hat{\bm{\theta}}) \ddot{\hat{\bm{x}}}_{obs} = &\ \mathrm{diag} (\hat{\bm{\lambda}}_{act} ) \bm{u} \\
    &- \bm{C}(\bm{q},\dot{\bm{x}},\hat{\bm{\theta}}) \dot{\hat{\bm{x}}}_{obs} + \bm{K}_{obs} \dot{\tilde{\bm{x}}}_{obs} 
\end{split}
\label{eq:adaptObs}
\end{equation}
where $\dot{\tilde{\bm{x}}}_{obs} = \dot{\bm{x}}-\dot{\hat{\bm{x}}}_{obs}$  is the generalized velocity observation error, and $\bm{K}_{obs} \in \mathbb{R}^{(6+n) \times (6+n)}$ is a positive definite gain matrix. The update laws for $\hat{\bm{\theta}}_i$  and  $\hat{\bm{\lambda}}_{act}$ are selected as
\begin{align}
\dot{\hat{\bm{\theta}}}_j &= - \mathop{\mathrm{Proj}}_{\bm{\theta}_j\in \mathcal{M}_c}\{ \gamma_{\theta_j} g(\hat{\bm{\theta}}_j)^{-1} \bm{Y}_j^T(\bm{q},\dot{\bm{x}},\dot{\hat{\bm{x}}}_{obs},\ddot{\hat{\bm{x}}}_{obs}) \dot{\tilde{\bm{x}}}_{obs}  \} \nonumber \\
% \label{eq:updateTheta} \\
\dot{\hat{\bm{\lambda}}}_{act} &=  \mathop{\mathrm{Proj}}_{\bm{\lambda}_{act} \in  \Lambda_{act}} \{ \bm{\Gamma}_\lambda \mathrm{diag} (\bm{u} ) \dot{\tilde{\bm{x}}}_{obs}  \}
\label{eq:updateThetaLambda}
\end{align}
where $\gamma_{\theta_i} \in \mathbb{R}$ is a positive scalar and $\bm{\Gamma}_\lambda  \in \mathbb{R}^{(6+n) \times (6+n)}$ is a positive definite gain matrix, 
\begin{equation*}
    \bm{Y}_j(\bm{q},\dot{\bm{x}},\dot{\hat{\bm{x}}}_{obs},\ddot{\hat{\bm{x}}}_{obs}) \bm{\theta}_j = \bm{Y}(\bm{q},\dot{\bm{x}},\dot{\hat{\bm{x}}}_{obs},\ddot{\hat{\bm{x}}}_{obs}) \bar{\bm{\theta}}_j\\
\end{equation*}
and $\bar{\bm{\theta}}_j= \left[ 0, \dots ,0, \bm{\theta}_j^T, 0, \dots 0 \right]^T \in \mathbb{R}^{10 \times (n+1)}$ has the same structure of $\bm{\theta}$ but with all the entries not related with $j^{th}$ rigid body set to zero. Finally, in the update law~\eqref{eq:updateThetaLambda},  $\mathrm{Proj}\{\cdot\}$ is the smooth projection operator defined in~\cite{khalil1996adaptive}, whereas  $g(\cdot) \in \mathbb{R}^{10 \times 10} $ is the pullback of the affine-invariant Riemannian metric on $\mathcal{P}(4)$ to $\mathbb{R}^{10}$ under the mapping $f(\cdot)$
\begin{equation}\label{eq:pullback}
    [g(\bm{\phi})]_{i,j}= \frac{1}{2} \mathrm{tr} \left(f(\bm{\phi})^{-1} f(e_i) f(\bm{\phi})^{-1} f(e_j) \right)
\end{equation}
where $e_i,e_j$ are vectors of the canonical basis of $\mathbb{R}^{10}$~\cite{lee2018geometric}. It should be noted that, since $\hat{\bm{\theta}}_j$ is projected into its feasible set $\mathcal{M}_{c}$, the matrix $g(\hat{\bm{\theta}}_j)$ is symmetric and positive definite, hence invertible. 
%
%
%
% that is a positive definite matrix and therefore always invertible, more details on this and on the projection operator are given in the appendices.
\begin{proposition}
\label{eq:prop1}
Assume that the forward trajectories $\dot{\bm{x}}(t)$, $\bm{u}(t)$ $t\geq 0$ exist for all $t\geq0$. Then, under the update law \eqref{eq:updateThetaLambda} all forward trajectories of the observer \eqref{eq:adaptObs} are bounded and satisfy:
\begin{equation}
 {\vert \vert \dot{\tilde{\bm{x}}}_{obs}(\cdot) \vert \vert}_{a}:=\limsup_{t \rightarrow \infty} {\vert \vert \dot{\tilde{\bm{x}}}_{obs}(t) \vert \vert}= 0
\end{equation}
being $\vert\vert \cdot \vert\vert_{a}$ the asymptotic norm of a signal defined in~\cite{teel1996nonlinear}.
\end{proposition}
%%Proof. \normalfont{
\begin{proof}
In the following, dependency of  $\bm{M}(\cdot)$, $\bm{C}(\cdot)$ on the exogenous signals $\bm{q}(t)$, $\dot{\bm{x}}(t)$, $\bm{u}(t)$ is  represented by a dependency on $t$. Consider the Lyapunov function candidate
\begin{equation}
    V(t,\dot{\tilde{\bm{x}}}_{obs},\tilde{\bm{\lambda}}_{act},\bm{\theta},\hat{\bm{\theta}}) =  V_e(t,\dot{\tilde{\bm{x}}}_{obs},\tilde{\bm{\lambda}}_{act}) + V_p(\hat{\bm{\theta}},\bm{\theta})
    \label{eq:lyapFunc}
\end{equation} 
where the first term
\begin{equation}
     V_e = \frac{1}{2} \dot{\tilde{\bm{x}}}_{obs}^T \bm{M}(t,\bm{\theta}) \dot{\tilde{\bm{x}}}_{obs}  + \frac{1}{2} \tilde{\bm{\lambda}}_{act}^T \bm{\Gamma}_{\lambda}^{-1} \tilde{\bm{\lambda}}_{act}
\end{equation}  
pertains to the observer error and the actuators efficiency,~and
\begin{equation}
    V_p = \sum_{j=0}^n \gamma_{\theta_j}^{-1} D_{F(\mathcal{P}(4))}(f(\hat{\bm{\theta}}_j) \vert \vert f(\bm{\theta}_j))
    \label{eq:lyapFuncVp}
\end{equation}
pertains to the inertial parameters. 
%With the exception of equation \eqref{eq:lyapFunc}, the augments of the various Lyapunov functions are omitted for compactness in the rest of the proof. 
Clearly, $V_e$ is positive definite, radially unbounded, and decrescent. In \cite{lee2018natural}, it is also proved that \eqref{eq:lyapFuncVp} is a suitable Lyapunov function candidate. To facilitate the next steps of the proof, note that
\begin{subequations}
\begin{align}
    \bm{M}(t,\bm{\theta})\ddot{\tilde{\bm{x}}}_{obs}  = & \ \bm{M}(t,\bm{\theta})\ddot{\bm{x}} -\bm{M}(t,\bm{\theta})\ddot{\hat{\bm{x}}}_{obs}\\
    \begin{split}
         =  &  \  \mathrm{diag} (\tilde{\bm{\lambda}}_{act} + \hat{\bm{\lambda}}_{act} )\bm{u}  - \bm{C}(t,\bm{\theta})\dot{\bm{x}}\\ 
        & \  -(\bm{M}(t,\tilde{\bm{\theta}})+\bm{M}(t,\hat{\bm{\theta}}) )\ddot{\hat{\bm{x}}}_{obs}
    \end{split} \\
    \begin{split}
     = & \ \mathrm{diag} (\bm{u})\tilde{\bm{\lambda}}_{act} - \bm{C}(t,\bm{\theta})\dot{\bm{x}} \\
    & \  -\bm{M}(t,\tilde{\bm{\theta}}) \ddot{\hat{\bm{x}}}_{obs} + \bm{C}(t,\hat{\bm{\theta}})\dot{\hat{\bm{x}}}_{obs}\\
    & \  - \bm{K}_{obs} \dot{\tilde{\bm{x}}}_{obs} 
    \label{eq:Mobserr_dd}
    \end{split}
\end{align}
\end{subequations}
where linearity of the operator $\mathrm{diag}(\cdot)$ operator and of the inertia matrix $\bm{M}(t,\cdot)$ in  the inertial parameters has been used together with~\eqref{eq:DynModEff} and~\eqref{eq:adaptObs}. The Lie derivative of $V_e$ is considered first:
\begin{subequations}
\begin{align}
    \begin{split}
     \dot{V_e} = & \ \dot{\tilde{\bm{x}}}_{obs}^T \bm{M}(t,\bm{\theta}) \ddot{\tilde{\bm{x}}}_{obs} + \frac{1}{2} \dot{\tilde{\bm{x}}}_{obs}^T \dot{\bm{M}}(t,\bm{\theta}) \dot{\tilde{\bm{x}}}_{obs} \\& + \dot{\tilde{\bm{\lambda}}}_{act}^T \bm{\Gamma}_{\lambda}^{-1} \tilde{\bm{\lambda}}_{act}
     \end{split}\\
     \begin{split}
     = & \ \dot{\tilde{\bm{x}}}_{obs}^T \left[  \mathrm{diag} (\bm{u})\tilde{\bm{\lambda}}_{act} - \bm{C}(t,\bm{\theta})(\dot{\hat{\bm{x}}}_{obs}+\dot{\tilde{\bm{x}}}_{obs}) \right. \\
     & \left. - \bm{M}(t,\tilde{\bm{\theta}}) \ddot{\hat{\bm{x}}}_{obs} + \bm{C}(t,\hat{\bm{\theta}}) \dot{\hat{\bm{x}}}_{obs} - \bm{K}_{obs} \dot{\tilde{\bm{x}}}_{obs}  \right]
     \\& + \frac{1}{2} \dot{\tilde{\bm{x}}}_{obs}^T \dot{\bm{M}}(t,\bm{\theta}) \dot{\tilde{\bm{x}}}_{obs} + \dot{\tilde{\bm{\lambda}}}_{act}^T \bm{\Gamma}_{\lambda}^{-1} \tilde{\bm{\lambda}}_{act} 
     \end{split}\\
     \begin{split}
     = & \ \dot{\tilde{\bm{x}}}_{obs}^T \mathrm{diag} (\bm{u})\tilde{\bm{\lambda}}_{act} + \dot{\tilde{\bm{\lambda}}}_{act}^T \bm{\Gamma}_{\lambda}^{-1}\tilde{\bm{\lambda}}_{act} 
     \\& \  - \dot{\tilde{\bm{x}}}_{obs}^T \bm{Y}(\bm{q},\dot{\bm{x}},\dot{\hat{\bm{x}}}_{obs},\ddot{\hat{\bm{x}}}_{obs}) \tilde{\bm{\theta}}  \\& \ -  \dot{\tilde{\bm{x}}}_{obs}^T \bm{K}_{obs} \dot{\tilde{\bm{x}}}_{obs}   
     \label{eq:Ve_dot3}
     \end{split}
\end{align}
\end{subequations}
where~\eqref{eq:Mobserr_dd} and the skew symmetry of $\dot{\bm{M}}(\cdot)-2\bm{C}(\cdot)$ have been used. Substituting the update law of $\hat{\bm{\lambda}}_{act}$ in~\eqref{eq:Ve_dot3}, the following inequality is obtained:
\begin{subequations}
\begin{align}
    \begin{split}
    \dot{V}_e \leq & -  \dot{\tilde{\bm{x}}}_{obs}^T \bm{K}_{obs} \dot{\tilde{\bm{x}}}_{obs} + \dot{\tilde{\bm{x}}}_{obs}^T \mathrm{diag} (\bm{u})\tilde{\bm{\lambda}}_{act}\\
    & \ - \mathop{\mathrm{Proj}}_{\bm{\lambda}_{act} \in  \Lambda_{act}} \{ \bm{\Gamma}_\lambda \mathrm{diag} (\bm{u} ) \dot{\tilde{\bm{x}}}_{obs}  \}^T \bm{\Gamma}_{\lambda}^{-1}\tilde{\bm{\lambda}}_{act} \\
    & \ - \dot{\tilde{\bm{x}}}_{obs}^T \bm{Y}(\bm{q},\dot{\bm{x}},\dot{\hat{\bm{x}}}_{obs},\ddot{\hat{\bm{x}}}_{obs}) \tilde{\bm{\theta}}
    \end{split}\\
     \leq & -  \dot{\tilde{\bm{x}}}_{obs}^T \bm{K}_{obs} \dot{\tilde{\bm{x}}}_{obs}  - \dot{\tilde{\bm{x}}}_{obs}^T \bm{Y}(\bm{q},\dot{\bm{x}},\dot{\hat{\bm{x}}}_{obs},\ddot{\hat{\bm{x}}}_{obs}) \tilde{\bm{\theta}} \label{eq:dotVe}
\end{align}
\end{subequations}
where the property $- \bm{\lambda}_{act}^T \mathrm{Proj}\{ \bm{\tau} \} \leq  - \bm{\lambda}_{act}^T \bm{\tau} $ of the projection operator have been used. Following~\cite{lee2018natural}, it is possible to show that the Lie derivative of $V_p$ satisfies
\begin{subequations}
\begin{align}
    \dot{V}_p = &\sum_{j=0}^n \gamma_{\theta_j}^{-1} \mathrm{tr} \left(f(\hat{\bm{\theta}}_j)^{-1} f(\dot{\hat{\bm{\theta}}}_j) f(\hat{\bm{\theta}}_j)^{-1} f(\tilde{\bm{\theta}}_j) \right) \label{eq:Vp_dot}\\
    = - &\sum_{j=0}^n \gamma_{\theta_j}^{-1} \dot{\hat{\bm{\theta}}}_j^T g(\hat{\bm{\theta}}_j) \tilde{\bm{\theta}}_j
    \label{eq:Vp_dot2}
\end{align}
\end{subequations}
owing to the properties of the pullback~\eqref{eq:pullback} (see~\cite{lee2018natural}) and the fact that \eqref{eq:Vp_dot} has the same structure of the inner product~\eqref{eq:innerProduct}. 
Substituting the update law of the inertial parameter estimates is \eqref{eq:Vp_dot2} and using the aforementioned property of the projection operator, one obtains
\begin{subequations}
\begin{align}
    \dot{V}_p = & \sum_{j=0}^n \dot{\tilde{\bm{x}}}_{obs}^T \bm{Y}_j(\bm{q},\dot{\bm{x}},\dot{\hat{\bm{x}}}_{obs},\ddot{\hat{\bm{x}}}_{obs}) \tilde{\bm{\theta}}_j \\
    = & \ \dot{\tilde{\bm{x}}}_{obs}^T \bm{Y}(\bm{q},\dot{\bm{x}},\dot{\hat{\bm{x}}}_{obs},\ddot{\hat{\bm{x}}}_{obs}) \tilde{\bm{\theta}} 
\end{align}\label{eq:dotVp}
\end{subequations}
%
%that is exactly the term depending on the inertial parameters identification error \eqref{eq:Ve_dot3} but with the opposite sign. 
Combining~\eqref{eq:dotVe} and~\eqref{eq:dotVp}, the Lie derivative of the Lyapunov function $V$ is found to satisfy the inequality
\begin{equation}
    \dot{V}  \leq -  \dot{\tilde{\bm{x}}}_{obs}^T \bm{K}_{obs} \dot{\tilde{\bm{x}}}_{obs} \leq 0
\end{equation}
Finally, the application of La Salle/Yoshizawa theorem yields boundedness of all trajectories of the adaptive observer and asymptotic convergence of the observation error of the generalized velocity. 
%}
\end{proof}

\subsection{Adaptive Controller}
In this subsection, a controller based on the observer dynamics is designed and convergence to zero of the tracking error is proven when $\dot{\tilde{\bm{x}}}_{obs}(t)=0$. To this end, the position and orientation errors of the BS are defined, along with the position error of the arm joints. Additionally, their corresponding velocity errors are also considered. Based on these quantities three augmented velocity errors are built: $\bm{v}_{err} \in \mathbb{R}^3$, $\bm{\omega}_{err}\in \mathbb{R}^3 $, $\dot{\bm{q}}_{err} \in \mathbb{R}^n$. The resulting errors dynamics having as inputs these new augmented velocities errors are proven to be input state stable (ISS). 
%meaning that if their inputs go to zero then the corresponding tracking errors will converge to zero. This characteristic is exploited in the design of the controller, which to ensure trajectory tracking, only needs to drive the augmented velocities errors to zero as a result.
The position and velocity error of the BS with respect to given smooth reference trajectories are defined as 
\begin{equation}
\begin{split}
    &\tilde{\bm{p}}_{b}^i = \bm{p}_{b}^i - \bm{p}_{b,ref}^i\\
    &\tilde{\bm{v}}_{b} = \bm{v}_{b} - \bm{v}_{b,ref} = \bm{R}_b^T \dot{\tilde{\bm{p}}}_{b}^i
    \label{eq:posVelErr}
\end{split}
\end{equation}
The orientation error and the associated angular velocity errors are defined as
\begin{equation}
\begin{split}
    &\tilde{\bm{R}} = \bm{R}_{b,ref}^T\bm{R}_{b}\\
    &\tilde{\bm{w}}_{b} = \bm{w}_{b} - \bm{w}_{b,ref} = \bm{R}_b^T \bm{w}_{b,ref}^{ref}
    \label{eq:orVelErr}
\end{split}
\end{equation}
To avoid manipulating rotation matrices, the orientation error  $\tilde{\bm{R}}\in SO(3)$  is parameterized via its Modified Rodriguez Parameters (MRP), $\tilde{\bm{\sigma}}\in \mathbb{R}^3$. Note that $\tilde{\bm{R}} = \mathbb{I}_{3 \times 3}$ corresponds to $\tilde{\bm{\sigma}}=0$. The propagation equation of the MRP is
\begin{equation}
    \dot{\tilde{\bm{\sigma}}} =  \frac{1}{2}\bm{G}(\tilde{\bm{\sigma}})\tilde{\bm{w}}_{b} 
    \label{eq:MRPdyn}
\end{equation}
where
\begin{equation*}
    \bm{G}(\tilde{\bm{\sigma}}):= \frac{1-\tilde{\bm{\sigma}}^T\tilde{\bm{\sigma}}}{2}\mathbb{I}_{3 \times 3} + \left[ \tilde{\bm{\sigma}} \right]_\times + \left[ \tilde{\bm{\sigma}} \right]_\times^2
\end{equation*}
The position and velocity errors of the joints are defined respectively as:
\begin{equation}
    \tilde{\bm{q}} =  \bm{q} - \bm{q}_{ref} \qquad \dot{\tilde{\bm{q}}} =  \dot{\bm{q}} - \dot{\bm{q}}_{ref} 
\end{equation}
The aforementioned augmented velocity errors are combined into a single vector $\dot{\bm{x}}_{err}\in \mathbb{R}^{6+n}$ defined as follows
\begin{equation}
    \underbrace{
    \begin{bmatrix}
        \bm{v}_{err}\\
        \bm{\omega}_{err}\\
        \dot{\bm{q}}_{err}\\
    \end{bmatrix}}_{\dot{\bm{x}}_{err}}=
    \underbrace{
    \begin{bmatrix}
        \bm{R}_b^T \bm{K}_p & \mathbb{0} & \mathbb{0}\\
        \mathbb{0} & \bm{K}_\sigma & \mathbb{0}\\
        \mathbb{0} & \mathbb{0} & \bm{K}_q\\
    \end{bmatrix}}_{\bm{K}_{\bar{x}}}
    \underbrace{
    \begin{bmatrix}
        \tilde{\bm{p}}_{b}^i\\
        \tilde{\bm{\sigma}}\\
        \tilde{\bm{q}}\\
    \end{bmatrix}}_{\bar{\bm{x}}}+
    \underbrace{
    \begin{bmatrix}
        \tilde{\bm{v}}_{b}\\
        \tilde{\bm{\omega}}_{b}\\
        \dot{\tilde{\bm{q}}}\\
    \end{bmatrix}}_{\dot{\tilde{\bm{x}}}} 
    \label{eq:augVelErr}
\end{equation}
where $\bm{K}_{\bar{x}}(\bm{R}_b) \in \mathbb{R}^{(6+n) \times (6+n)}$, $\bm{K}_{p} \in \mathbb{R}^{3 \times 3}$, $\bm{K}_{\sigma} \in \mathbb{R}^{3 \times 3}$, $\bm{K}_{q} \in \mathbb{R}^{n \times n}$ are positive definite matrices, $\bar{\bm{x}} \in \mathbb{R}^{(6+n)}$ is the modified position/orientation error, and $\dot{\tilde{\bm{x}}} = \dot{\bm{x}} - \dot{\bm{x}}_{ref} \in \mathbb{R}^{(6+n)}$ is the velocity tracking error. 

\begin{proposition}
\label{eq:propISSerr}
The dynamics of the modified position/orientation error $\bar{\bm{x}}$ is ISS w.r.t. the input $\dot{\bm{x}}_{err}$. In particular, the asymptotic bounds
\begin{subequations}
\begin{align}
    &{\vert \vert \tilde{\bm{p}}_{b}^i(t) \vert \vert}_a \leq   \frac{1}{\lambda_{min}(\bm{K}_{p})} {\vert \vert \bm{v}_{err}(t) \vert \vert}_a\\
    & {\vert \vert \tilde{\bm{\sigma}}(t) \vert \vert}_a \leq   \frac{1}{\lambda_{min}(\bm{K}_{\sigma})} {\vert \vert \bm{w}_{err}(t) \vert \vert}_a\\
    & {\vert \vert \tilde{\bm{q}}(t) \vert \vert}_a \leq \frac{1}{\lambda_{min}(\bm{K}_{q})} {\vert \vert \bm{q}_{err}(t) \vert \vert}_a
\end{align}
\end{subequations}
hold, where $\lambda_{min}(\cdot)$ denotes the minimum eigenvalue of a matrix. 
\end{proposition}
%and
%%
%\begin{equation*}
%    {\vert \vert \cdot \vert \vert}_a = \limsup_{t \rightarrow \infty} {\vert \vert \cdot \vert \vert} 
%\end{equation*}
%%
%\noindent
%Proof. \normalfont{
\begin{proof}
The result follows immediately from~\cite[Lemma 3.3]{teel1996nonlinear} applied to the Lyapunov functions $V(\tilde{\bm{p}}_{b}^i) = \left(\tilde{\bm{p}}_{b}^i\right)^T\tilde{\bm{p}}_{b}^i$, $ V(\tilde{\bm{\sigma}}) = 2 \ln(1 + \tilde{\bm{\sigma}}^T \tilde{\bm{\sigma}})$, and $V(\tilde{\bm{q}}) = \tilde{\bm{q}}^T\tilde{\bm{q}}$, respectively. 
%\qed
%}
\end{proof}
The remaining objective is to regulate $\dot{\bm{x}}_{err}(t)$ to zero. 
%To achieve this its dynamics are first analyzed. 
To this end, consider the dynamics of \eqref{eq:augVelErr}
\begin{equation}
    \ddot{\bm{x}}_{err} = \dot{\bm{K}}_{\bar{x}}(\bm{R}_b,\bm{\omega}_b) \bar{\bm{x}} + \bm{K}_{\bar{x}}(\bm{R}_b) \dot{\bar{\bm{x}}} + \ddot{\tilde{\bm{x}}}
    \label{eq:augAccErr} 
\end{equation}
where $\dot{\bar{\bm{x}}} \in \mathbb{R}^{6+n}$, $\dot{\bm{K}}_{\bar{x}}(\bm{R}_b,\bm{\omega}_b) \in \mathbb{R}^{(6+n) \times (6+n)}$  can be easily derived from equations~\eqref{eq:posVelErr}, \eqref{eq:orVelErr} and \eqref{eq:MRPdyn}.
 Substituting  $ \ddot{\bm{x}} = \ddot{\tilde{\bm{x}}} + \ddot{\bm{x}}_{ref}$ in~\eqref{eq:DynModEff} and using~\eqref{eq:augAccErr}, the dynamics of $\dot{\bm{x}}_{err}$ is written as
\begin{equation}
\begin{split}
    \bm{M}(t,\bm{\theta}) \ddot{\bm{x}}_{err} = & - \bm{C}(t,\bm{\theta})  \dot{\bm{x}} + \bm{M}(t,\bm{\theta}) \Big( \bm{K}_{\bar{x}}(\bm{R}_b) \dot{\bar{\bm{x}}}  \\
     & \left. + \dot{\bm{K}}_{\bar{x}}(\bm{R}_b,\bm{\omega}_b) \bar{\bm{x}} \right)\\
     &- \bm{M}(t,\bm{\theta}) \ddot{\bm{x}}_{ref} + \mathrm{diag} (\bm{\lambda}_{act} ) \bm{u} 
\end{split}
\end{equation}
It is clear from the formula above that the $\ddot{\bm{x}}_{err}$ dynamics depend on the unknown parameter $\bm{\theta}$. To overcome this problem one needs only define an estimated version of the augmented velocity errors vector $\dot{\hat{\bm{x}}}_{err}$ and to control its dynamics, which depends on  the adaptive observer \eqref{eq:adaptObs}. To this end, in reference to equation~\eqref{eq:augVelErr}, the estimated augmented error $\dot{\hat{\bm{x}}}_{err}$ is defined as follows: 
\begin{subequations}
\begin{align}
    \dot{\bm{x}}_{err} & = \bm{K}_{\bar{x}}(\bm{R}_b,\bm{\omega}_b) \bar{\bm{x}} + \dot{\tilde{\bm{x}}}\\
    \label{eq:estAugTrack1}
    & = \bm{K}_{\bar{x}}(\bm{R}_b,\bm{\omega}_b) \bar{\bm{x}} + \underbrace{\dot{\bm{x}} - \dot{\hat{\bm{x}}}_{obs}}_{\dot{\tilde{\bm{x}}}_{obs}} + \underbrace{\dot{\hat{\bm{x}}}_{obs} - \dot{\bm{x}}_{ref}}_{\dot{\bm{e}}_{\dot{x}}}\\
    & = \dot{\hat{\bm{x}}}_{err} + \dot{\tilde{\bm{x}}}_{obs}
    \label{eq:estAugTrack3}
\end{align}
\end{subequations}
where $\dot{\bm{e}}_{\dot{x}}\in \mathbb{R}^{6+n}$ is the estimated tracking error. From \eqref{eq:estAugTrack3} it follows that asymptotic regulation of  $\dot{\tilde{\bm{x}}}_{obs}(t)$ implies regulation of  $\dot{\hat{\bm{x}}}_{err}(t)- \dot{\bm{x}}_{err}(t)$, and the former event is ensured by the properties of the observer established in Proposition~\ref{eq:prop1}.
The dynamics of $\dot{\hat{\bm{x}}}_{err}$ are obtained from the derivative of~\eqref{eq:estAugTrack1}
\begin{equation}
    \ddot{\hat{\bm{x}}}_{err} = \dot{\bm{K}}_{\bar{x}}(\bm{R}_b,\bm{\omega}_b) \bar{\bm{x}} + \bm{K}_{\bar{x}}(\bm{R}_b) \dot{\bar{\bm{x}}} + \ddot{\hat{\bm{x}}}_{obs} - \ddot{\bm{x}}_{ref}
\end{equation}
Solving for $\ddot{\hat{\bm{x}}}_{obs} $ in the above identity and substituting the result in the equation of the observer dynamics~\eqref{eq:adaptObs} yields the dynamics of $\dot{\hat{\bm{x}}}_{err}$ in the form
\begin{equation*}
\begin{split}
    \bm{M}(t,\hat{\bm{\theta}}) \ddot{\hat{\bm{x}}}_{err} = & - \bm{C}(t,\hat{\bm{\theta}})  \dot{\hat{\bm{x}}}_{obs}  + \bm{M}(t,\hat{\bm{\theta}}) \left( \bm{K}_{\bar{x}}(\bm{R}_b) \dot{\bar{\bm{x}}} \right. \\
     & \left. + \dot{\bm{K}}_{\bar{x}}(\bm{R}_b,\bm{\omega}_b) \bar{\bm{x}} \right) - \bm{K}_{obs} \dot{\tilde{\bm{x}}}_{obs}\\
     &- \bm{M}(t,\hat{\bm{\theta}}) \ddot{\bm{x}}_{ref} + \mathrm{diag} (\hat{\bm{\lambda}}_{act} ) \bm{u} 
\end{split}
\end{equation*}
Let $\tilde{\bm{u}}:= \bm{u}-\bm{u}_{c}$ be the mismatch between the actual forces and torques generated by the actuators and the commanded ones to the actuators by the control system. The control input
\begin{equation*}
\begin{split}
    \mathrm{diag} (\hat{\bm{\lambda}}_{act} ) \bm{u}_c\, := & \,\bm{C}(t,\hat{\bm{\theta}}) \dot{\hat{\bm{x}}}_{obs}  - \bm{M}(t,\hat{\bm{\theta}}) \left( \bm{K}_{\bar{x}}(\bm{R}_b) \dot{\bar{\bm{x}}} \right. \\
     & \left. + \dot{\bm{K}}_{\bar{x}}(\bm{R}_b,\bm{\omega}_b) \bar{\bm{x}} \right) - \bm{K}_{obs} \dot{\bm{x}}_{err}\\
     &+ \bm{M}(t,\hat{\bm{\theta}}) \ddot{\bm{x}}_{ref} 
\end{split}
\end{equation*}
yields the closed-loop dynamics
\begin{equation}
    \bm{M}(t,\hat{\bm{\theta}}) \ddot{\hat{\bm{x}}}_{err} = - \bm{K}_{obs} \dot{\hat{\bm{x}}}_{err} + \mathrm{diag} (\hat{\bm{\lambda}}_{act} ) \tilde{\bm{u}} 
    \label{eq:xerrHatDyn}
\end{equation}

\begin{proposition}\label{prop:ISSerr}
\label{eq:prop2}
Let assumptions of Proposition \ref{eq:prop1} hold and assume that $\tilde{\bm{u}}(t), \ t>0$ is bounded. Then the system \eqref{eq:xerrHatDyn} is ISS w.r.t. the input $\tilde{\bm{u}}$ and satisfies the asymptotic bounds:
\begin{equation}
\hspace{-2mm}
    {\vert \vert \dot{\hat{\bm{x}}}_{err}(t) \vert \vert}_a \leq  \frac{\lambda_{max}\left(\bm{M}(\bm{q}^*,\hat{\bm{\theta}}^*)\right)}{\lambda_{min}(\bm{K}_{obs})} {\vert \vert \tilde{\bm{u}}(t) \vert \vert}_a\\ 
\end{equation}
\noindent
where $\lambda_{max}(\cdot)$ represents the maximum eigenvalue of the argument matrix and $\bm{q}^* \in (-\pi,\pi]^n,\hat{\bm{\theta}}^* \in \mathcal{M}_c^n $ are respectively the joints configuration and inertial parameters that lead to the bigger eigenvalue of $\bm{M}(\cdot)$.
\end{proposition}
\begin{proof}
The result follows directly from~\cite[Lemma 3.3]{teel1996nonlinear} applied to the Lyapunov function $V(\dot{\hat{\bm{x}}}_{err}) = \dot{\hat{\bm{x}}}_{err}^T \dot{\hat{\bm{x}}}_{err}$. 
\end{proof}
The result of Proposition~\ref{prop:ISSerr} establishes robustness of the adaptive controller with respect to bounded mismatches between the actual and the commanded control input, that is, robustness to bounded input perturbations. Note that asymptotic regulation of the estimated augmented error (hence the tracking error) is recovered for vanishing input disturbances. 
\section{Simulation Study}\label{sec:Sim}
The effectiveness of the control solution proposed in this work is analyzed in simulations. A challenging scenario is considered in order to test the performance and the robusteness of the controller against model uncertainty, actuators wear and unmodelled actuator nonlinearities.  
\subsection{Simulation Setup}
{\renewcommand{\arraystretch}{1.3}
 
\begin{table}[t!]
\begin{minipage}{\linewidth}
    \centering
    \caption{Chaser Satellite Characteristics}
    \begin{tabular}{|c|c|c|}
        \hline
        Mass & Position of the COM & Inertia ($\bm{I}_b$)\\
        \hline
        \multicolumn{1}{|c|}{\multirow{4}{*}{$m_b = 1900 \ \mathrm{kg}$}} & \multicolumn{1}{|c|}{\multirow{4}{*}{$\bm{p}_{bj} = \bm{0} \ \mathrm{m}$}} & $I_{xx} = 13500 \ \mathrm{kg\, m}^2$\\
        & & $I_{yy} = 2000 \ \mathrm{kg\, m}^2$\\ 
        & & $I_{zz} = 14000 \ \mathrm{kg\, m}^2$\\ 
        & & $I_{ij} = 0 \ \mathrm{kg\, m}^2$\\
        \hline        
    \end{tabular}
    \label{tab:BScha}
     \vspace*{0.5 cm}
\end{minipage}
\\
\begin{minipage}{\linewidth}
    \centering
    \caption{Arm's Denavit-Hartenberg parameters (See. \cite{Siciliano})}
    \begin{tabular}{|c|c|c|c|c|c|c|c|}
    \hline
     & $l_1$ & $l_2$ & $l_3$ & $l_4$ & $l_5$ & $l_6$ & $l_7$ \\
    \hline
    $a_i \ [{\rm m}]$ & 0 & 0 & 0 & 0 & 0 & 0 & 0 \\
    \hline
    $\alpha_i \ [\mathrm{rad}]$ & $\pi/2$ & $-\pi/2$& $\pi/2$& $-\pi/2$& $\pi/2$
    &$-\pi/2$ & 0 \\
    \hline
    $d_i \ [{\rm m}]$ & 0.3 & 0.16 & 1.15 & -0.16 & 1.15 & -0.16 & 0.4 \\
    \hline
    \end{tabular}
    \vspace*{0.5 cm}
    \label{tab:DH}
\end{minipage}
\\
\begin{minipage}{\linewidth}
    \centering
    \caption{Controller Parameters}
    \begin{tabular}{|c|}
        \hline
         Parameters\\
        \hline
        $\bm{K}_{obs} = 2.5 \ \mathbb{I}_{13 \times 13},\ \bm{K}_{p} = 0.2 \ \mathbb{I}_{3 \times 3},\ \bm{K}_{\sigma} = 0.2 \ \mathbb{I}_{3 \times 3},$\\
        $\bm{K}_{q} = 0.2 \ \mathbb{I}_{7 \times 7},\ \gamma_{\theta_j} = 20,$\\
        $\bm{\Gamma}_{\lambda} =  2 \ \mathbb{I}_{13 \times 13}, \delta = 10^{-3},\ \lambda_{min}=0.1$\\
        %$ a = [1,-20,-20,-200,0.1,0.1,0.1,-10,-10,-10]^T $\\
        %$ b = [200,20,20,200,100,100,100,10,10,10]^T $\\
        \hline
    \end{tabular}
    \vspace*{0.5 cm}
    \label{tab:contrParam}
\end{minipage}
\\
\begin{minipage}{\linewidth}
    \centering
    \caption{Grasped Object Characteristics}
    \begin{tabular}{|c|c|c|c|}
        \hline
        & Mass & Pos. COM & Inertia\\
        \hline
        \multicolumn{1}{|c|}{\multirow{3}{*}{A Priori Par.}} & \multicolumn{1}{c|}{\multirow{3}{*}{$m = 30 \ \mathrm{kg}$}} & $\bm{h}_{c}^x = 0 \ \mathrm{kg\,m}$ & \multicolumn{1}{c|}{\multirow{3}{*}{\shortstack[l]{$I_{ii} = 40 \ \mathrm{kg\,m}^2$ \\ $I_{ij} = 0 \ \mathrm{kg\,m}^2$}}} \\
        & & $\bm{h}_{c}^y = 0 \ \mathrm{kg\,m}$ & \\ 
        & & $\bm{h}_{c}^z = 12 \ \mathrm{kg\,m}$ &\\ 
        \hline  
        \multicolumn{1}{|c|}{\multirow{4}{*}{True Par.}} & \multicolumn{1}{c|}{\multirow{4}{*}{$m = 100 \ \mathrm{kg}$}} & \multicolumn{1}{c|}{\multirow{4}{*}{\shortstack[l]{$\bm{h}_{c}^x = 0 \ \mathrm{kg\,m}$ \\ $\bm{h}_{c}^y = 0 \ \mathrm{kg\,m}$ \\ $\bm{h}_{c}^z = 40 \ \mathrm{kg\,m}$}}} & $I_{xx} = 80 \ \mathrm{kg\,m}^2$\\
        & & & $I_{yy} = 75 \ \mathrm{kg\,m}^2$\\ 
        & & & $I_{zz} = 90 \ \mathrm{kg\,m}^2$ \\ 
        & & & $I_{ij} = 0 \ \mathrm{kg\,m}^2$ \\ 
        \hline 
    \end{tabular}
    \label{tab:goInert}
\end{minipage}
\end{table}}

Simulations are performed on a multibody model of the system build in Matlab Simulink using the Mechanics toolbox. The characteristics of the SMS are inspired by actual space missions. The BS is represented as a box-shaped rigid body with two solar panels attached to its sides. Its inertial parameters are listed in Table~\ref{tab:BScha}. The BS actuation system consists of a RCS and a group of four RWs whose dynamics are simulated in detail. The RCS is made up of $24$ jet thrusters that together can generate a thrust of $40$~[N] and a torque of $40$~[Nm] along each axis. The RWs together are able to generate $0.5$~[Nm] of torque along each axis. The structure of the robotic arm is identical to the one described in reference \cite{armStruct}. It consists of seven links connected by revolute joints. The Denavit-Hartenberg parameters (see \cite{Siciliano}) of this configuration are listed in Table~\ref{tab:DH}. The links are assumed to be hollow aluminum cylinders with a thickness of $13.5$~[mm] and an external radius of $63.5$~[mm]. The dynamics of the joints motors are also simulated and it is assumed that they can generate a maximum torque of $10$~[Nm]. The controller parameters used in the simulation are reported in Table~\ref{tab:contrParam}. 

In the first part of the simulation, i.e., within the interval $[5,35]$~[s], the BS is moved diagonally while keeping the position of arm joints fixed. Then, between $t=70$~[s] and $t=170$~[s], an eight-shaped reference trajectory with a non-smooth velocity profile at its two ends is imposed on the EE while the pose of the BS remains fixed. At the beginning of the simulation, the BS pose and the position of two joints are set to be misaligned with the reference trajectory. Specifically, the BS is misaligned by $0.1$~[m] in the direction of the $x$-axis, using the $xyz$ Euler Angle representation, by $[0,\pi/8,\pi/8]_{xyz}$~[rad]. In addition, also the fourth and sixth joints are initially misaligned with respect to the reference trajectory by $-\pi/6$~[rad] and $+\pi/6$~[rad], respectively.
Furthermore, the a priori information about the grasped object inertial parameters is incorrect (see Tab. \ref{tab:goInert}), and an efficiency loss of two joint motors is scheduled to occur in the middle of the simulation.  In particular, at $t=120$~[s] the efficiency of the first and fourth joints motors is reduced to $70$~\% and $80$~\% of the full efficiency, respectively.

\subsection{Simulation Results}
\begin{figure}[t!]
    \begin{minipage}{0.91\linewidth}
        \centering
        \includegraphics[width=1\textwidth]{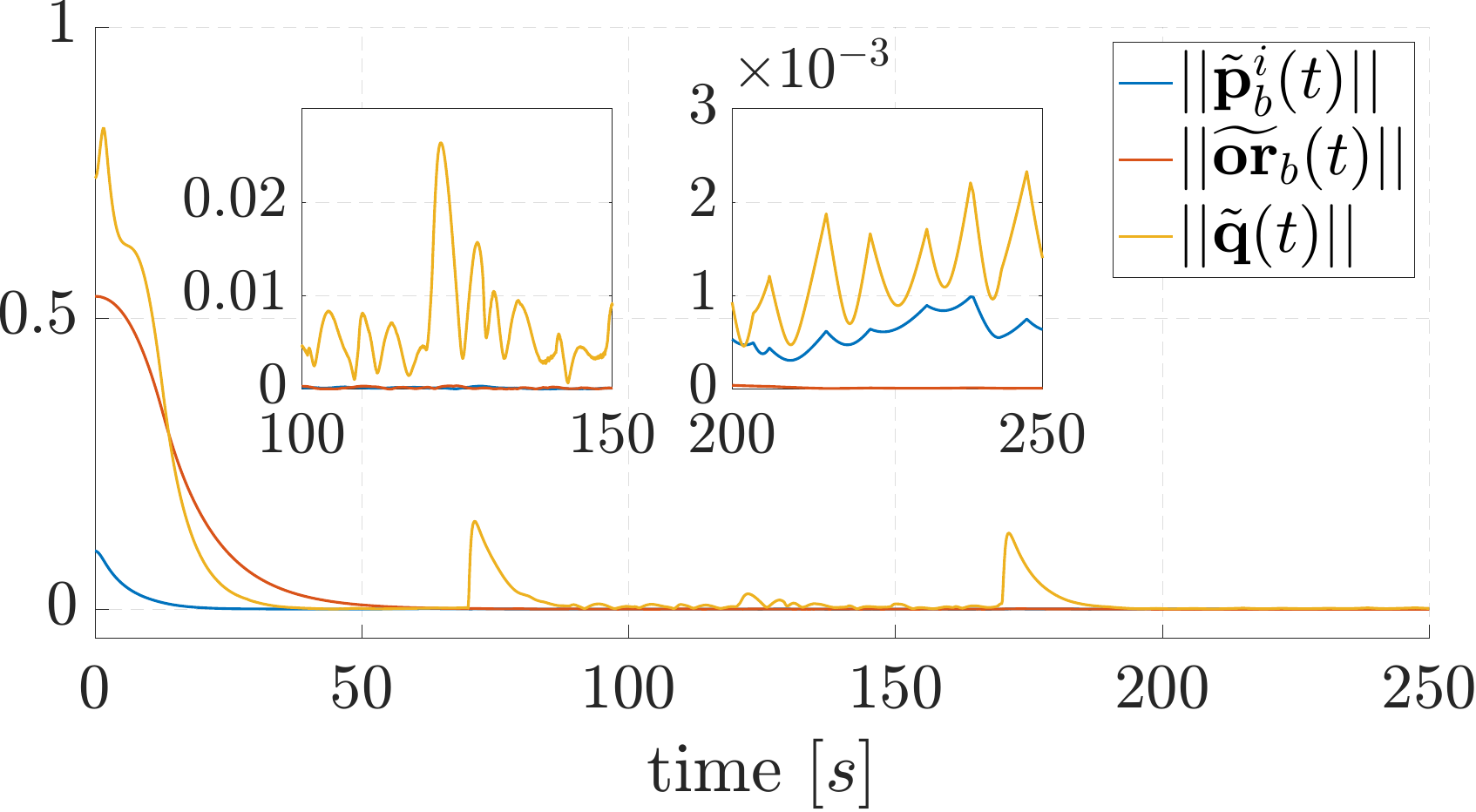}
        \caption{ Norm of the BS position error, $\tilde{\bm{p}}_{b}^{i}$;  norm of the BS attitude error (expressed in Euler Angles), $\widetilde{\bm{or}}_{b}$;  and norm of the joint position error,~$\tilde{\bm{q}}$.}
        \label{fig:Traj_Err_Norm}
    \end{minipage}
    \vspace*{0.5 cm}
    \\
    \begin{minipage}{0.91\linewidth}
        \centering
        \centering
        \includegraphics[width=1\textwidth]{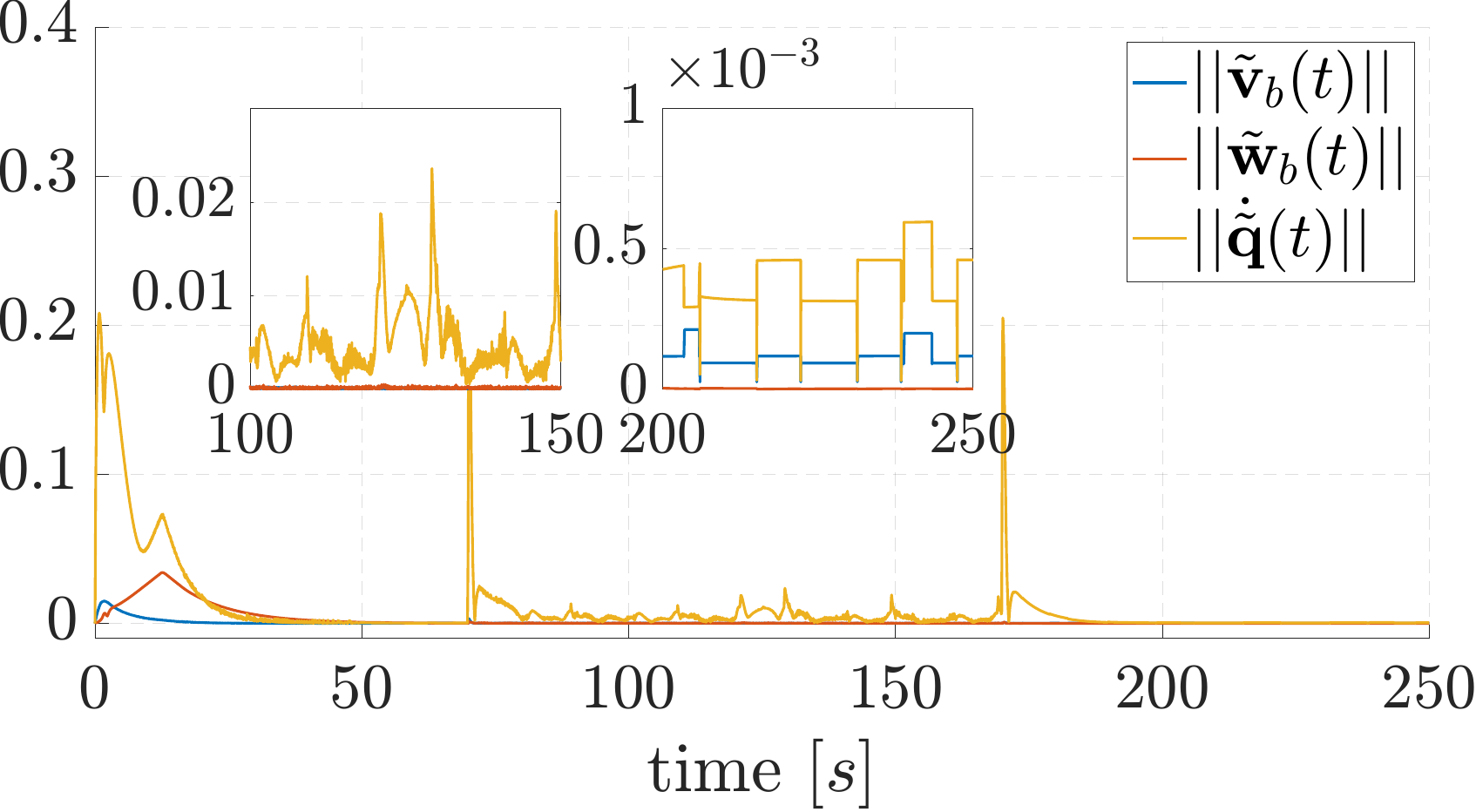}
        \caption{Norm of the BS velocity error, $\tilde{\bm{v}}_{b}$; norm of the  BS angular velocity error, $\widetilde{\bm{w}}_{b}$; and norm of the joint velocity error, $\dot{\tilde{\bm{q}}}$.}
        \label{fig:Traj_Vel_Err_Norm}
    \end{minipage}
\end{figure}

%In the following, the performance of the controller during the aforementioned scenario is commented and analyzed. 
The results of the simulations are first presented in Fig.~\ref{fig:Traj_Err_Norm}, which shows the evolution of the tracking error. In the figure, each of the above-mentioned phases of the simulation can be easily recognized. 
Excluding the first $70$~s when the SMS converges to the desired trajectory, three peaks are noticeable in the arm joint position error. The first and last ones can be attributed to the jumps in the velocity reference, whereas the middle one can be attributed to the efficiency loss of the two arm joints. Despite these challenges, the presence of uncertainties, and the nonlinearities of the actuation system, it is possible to notice how the errors are quickly controlled to a neighborhood of zero. Keeping in mind that the force produced by the RCS can be either zero or above a minimum threshold, which depends on the fuel valves opening times, smaller error values could be archived at the price of a higher control effort. Similar conclusions can be inferred from the velocities error norms represented in Fig.~\ref{fig:Traj_Vel_Err_Norm}. In addition, from the same image, the effects of the lower bound on the force generated by the RCS can be appreciated by observing the discrete-like values of the errors assumed in the interval $[200,250]$~[s].  

\begin{figure}[t!]
    \begin{minipage}{0.88\linewidth}
        \centering
        \includegraphics[width=1\textwidth]{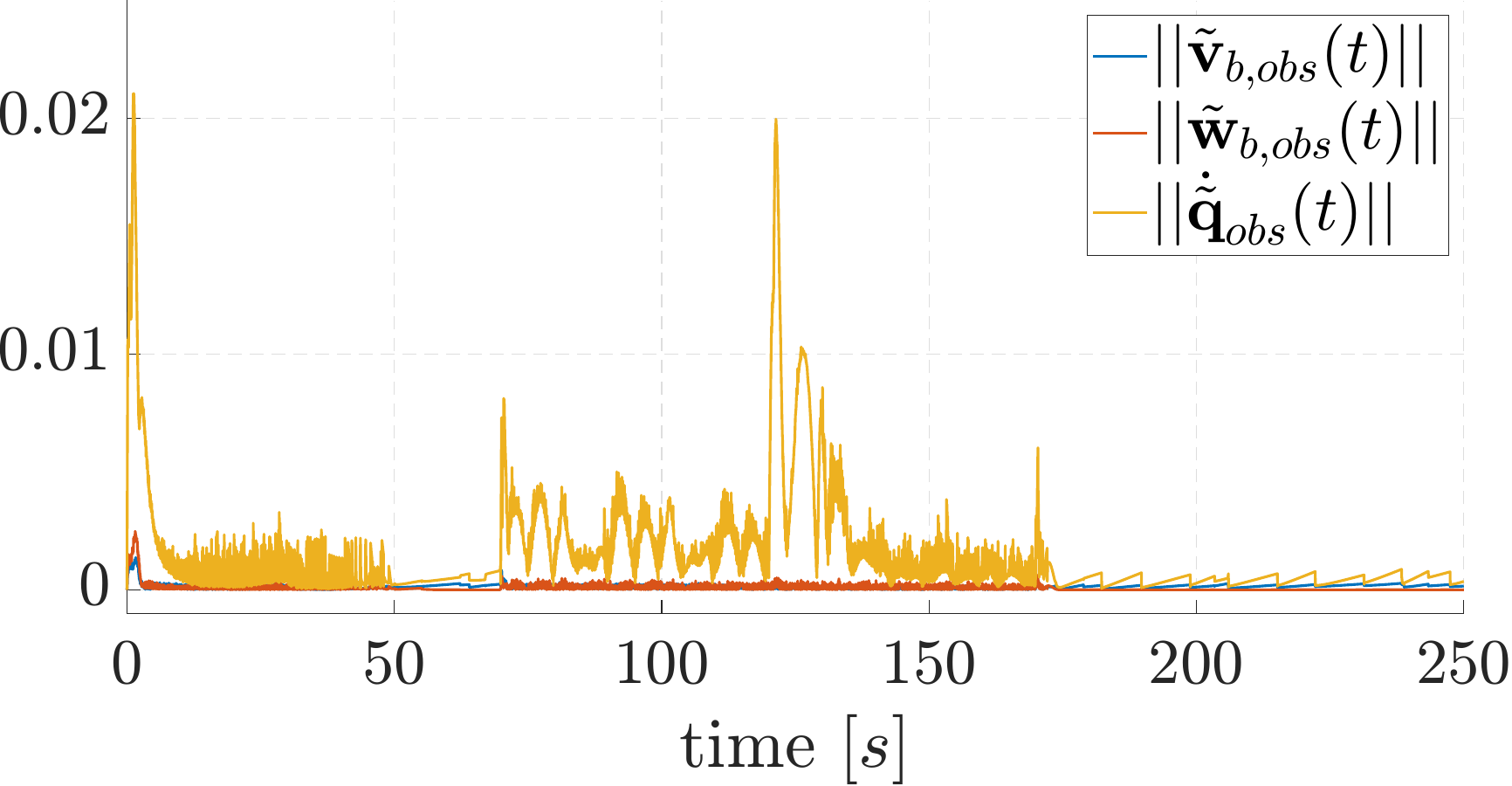}
        \caption{Norm of the BS observation velocity error, $\tilde{\bm{v}}_{b,obs}$; norm of the  BS observation angular velocity error, $\widetilde{\bm{w}}_{b,obs}$; and norm of the joint observation velocity error, $\dot{\tilde{\bm{q}}}_{obs}$.}
        \label{fig:Obs_Err}
    \end{minipage}
    \vspace*{0.5 cm}
    \\
    \begin{minipage}{0.92\linewidth}
        \centering
        \includegraphics[width=1\textwidth]{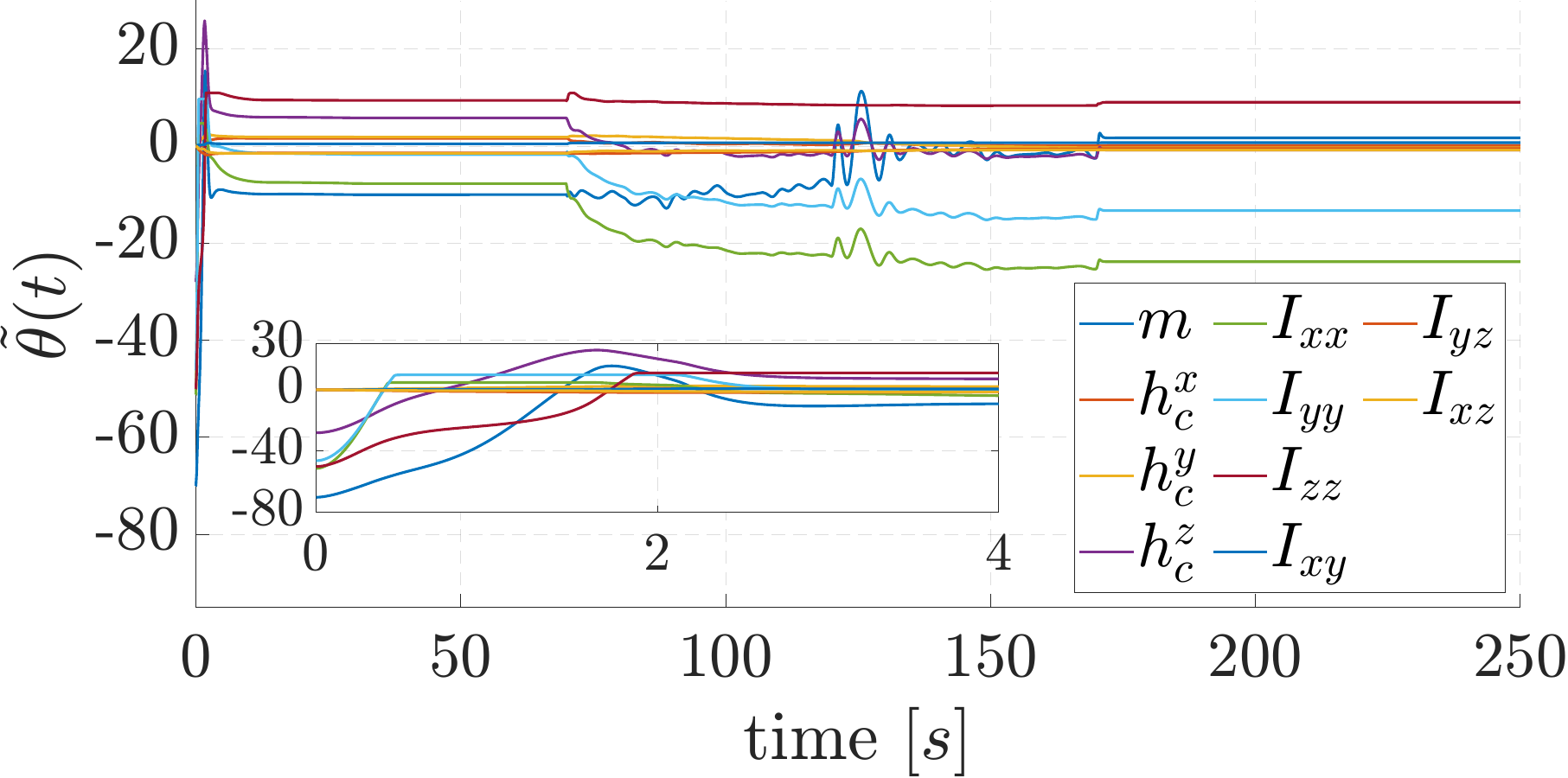}
        \caption{Estimation error of the inertial parameters of the combined grasped object and EE.}
        \label{fig:pi_vec}
    \end{minipage}
    \vspace*{0.5 cm}
    \\
    \begin{minipage}{0.92\linewidth}
        \centering
        \centering
        \includegraphics[width=1\textwidth]{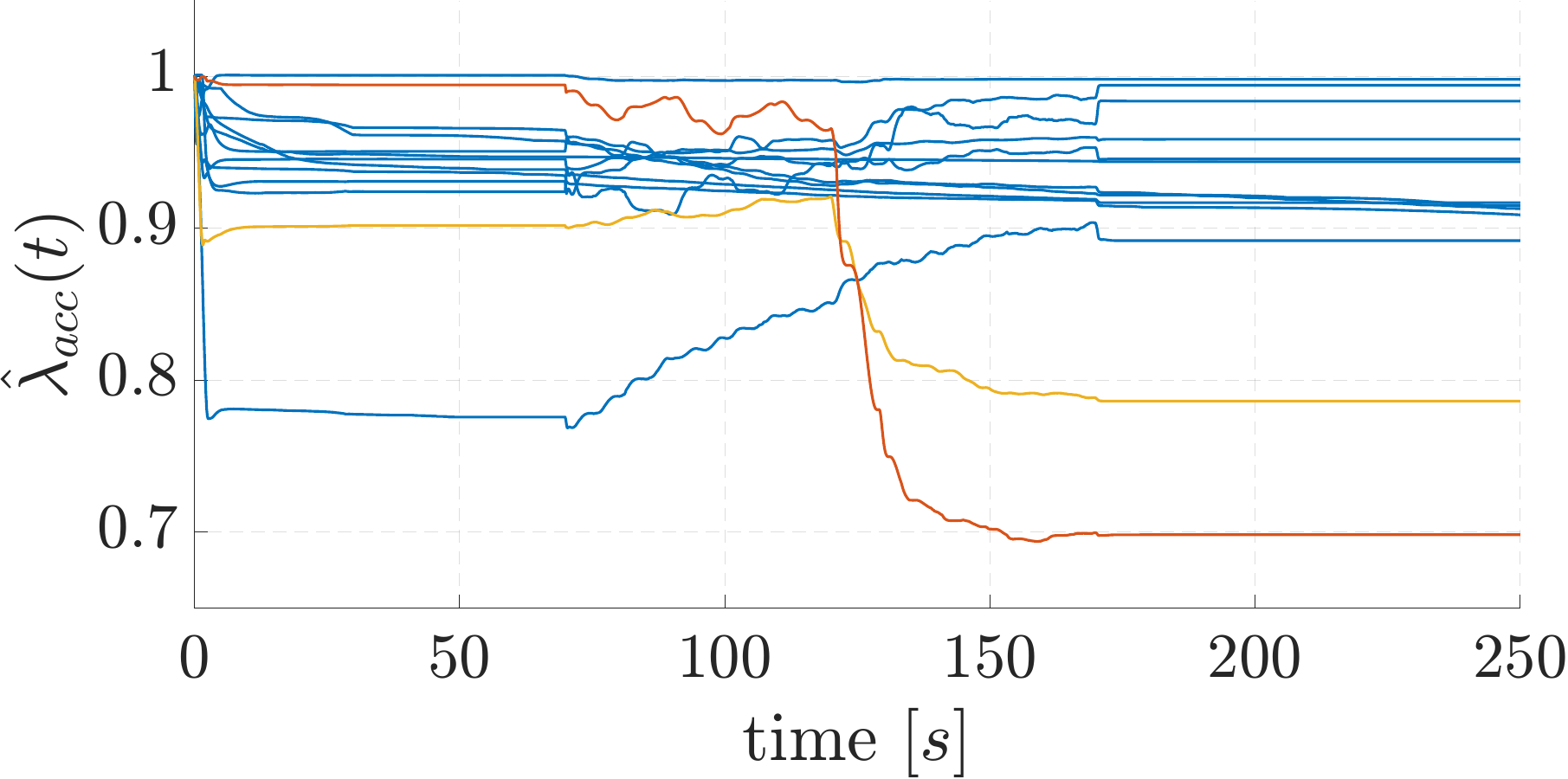}
        \caption{ Estimated actuators efficiencies. The two joints whose efficiency is reduced at $t=120$~[s] are highlighted in red and yellow, the others are in blue.}
        \label{fig:lambda}
    \end{minipage}
\end{figure}
Additional considerations can be deduced by looking at the norm of the observation errors, the estimation error of the inertial parameters of the EE, and the estimated actuator efficiency, reported in Fig.~\ref{fig:Traj_Vel_Err_Norm}, Fig.~\ref{fig:pi_vec} and Fig.~\ref{fig:lambda} respectively. 
In the first few seconds, the adaptive parameters of the observer are tuned until the observation error is almost zero. The adaptation slows down until the EE is moved when the dynamics are excited. It is worth noticing that the partial failure of the two joint motors helps in reducing the observation error after a transient. This behavior can be explained by the fact that new dynamics are again excited by the malfunctions, leading to the excitation of the regressor in the update law. The norm of the observation error of joints velocities exhibits a noisy behavior, due to the fact that simple saturation and dead zone effects for the RCS are included in the actuator model used in the observer, while the actual RCS dynamics are simulated using realistic on-off mode and pulse modulation effects.

\begin{figure}[t!]
    \begin{minipage}{0.92\linewidth}
        \centering
        \includegraphics[width=1\textwidth]{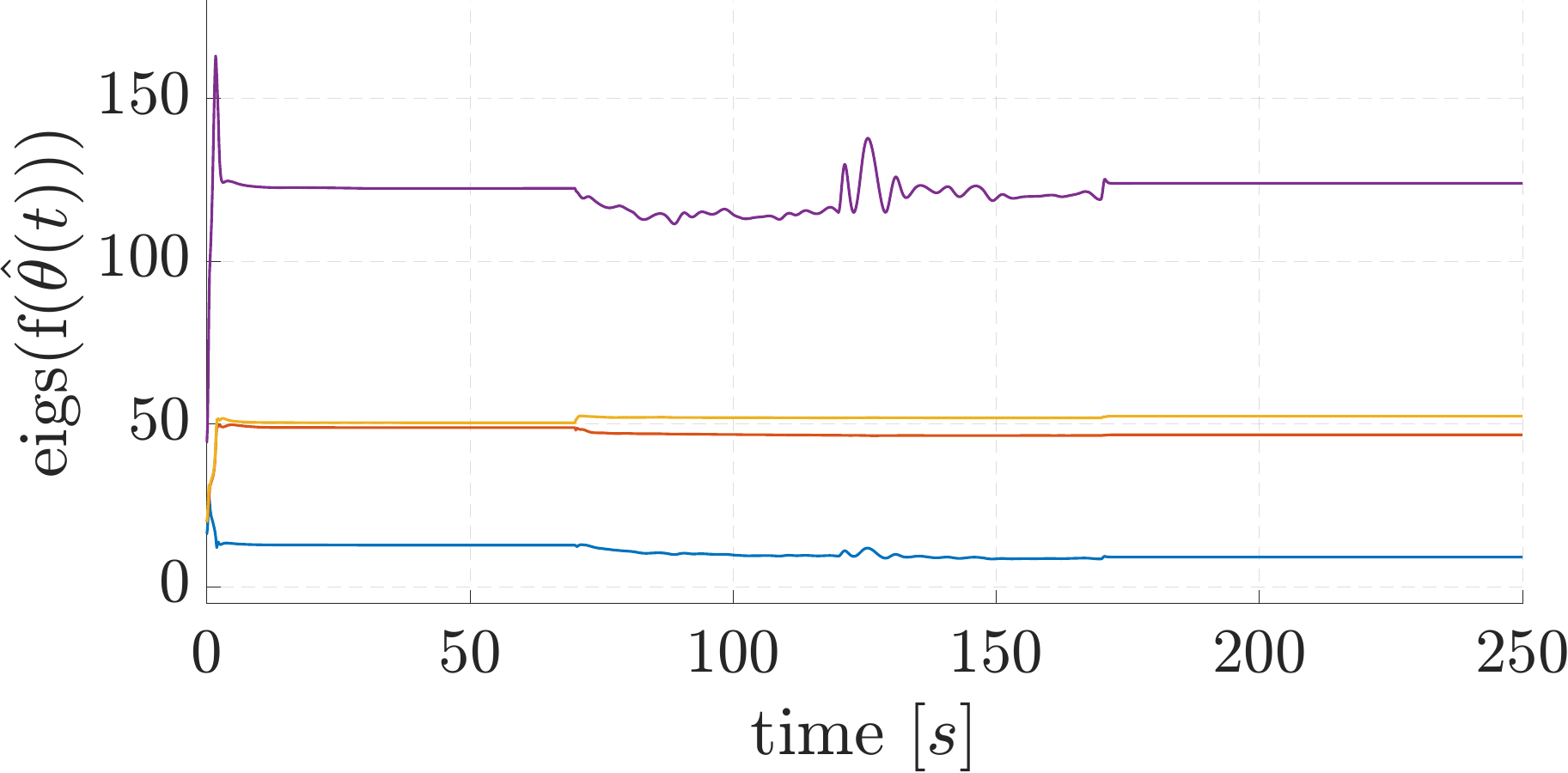}
        \caption{\footnotesize Eigenvalues of the estimated inertial parameters of the combined grasped object and EE, mapped on $\mathcal{P}(4)$.}
        \label{fig:eigs}
    \end{minipage}
\end{figure}

Examination of Fig.~\ref{fig:eigs}, which shows the time history of the eigenvalues of $f(\hat{\bm{\theta}})$, confirms that across the simulation the estimated inertial parameters of the EE are always physically consistent. 

\section{Conclusions}
\label{sec:Conc}

In this work, an indirect adaptive-based control solution is proposed to control a SMS in the presence of uncertainties and disturbances. First, an observer for the generalized velocities of the system is designed with the purpose of estimating online a set of parameters that capture the stimulated system dynamics. Then, a controller is designed on the basis of the adaptive observer serving as a proxy of the actual system dynamics. The performance of both the observer and controller is analyzed. Finally, a high-fidelity simulation environment, where realistic dynamics of both the system and the actuators are modeled, is employed to test closed-loop performance. A partially known grasped object is moved along a non-smooth trajectory during which an efficiency loss of the actuators is simulated. Notwithstanding the presence of model uncertainties and disturbances, the controller is capable of achieving remarkable convergence performance of the tracking error and boundedness of all closed-loop trajectories, while preserving the physical consistency of the estimated inertial parameters.

\bstctlcite{IEEEexample:BSTcontrol}
\bibliographystyle{IEEEtran}
\bibliography{ACC2024-adaptiveSMS}

\begin{thebibliography}{10}
\providecommand{\url}[1]{#1}
\csname url@samestyle\endcsname
\providecommand{\newblock}{\relax}
\providecommand{\bibinfo}[2]{#2}
\providecommand{\BIBentrySTDinterwordspacing}{\spaceskip=0pt\relax}
\providecommand{\BIBentryALTinterwordstretchfactor}{4}
\providecommand{\BIBentryALTinterwordspacing}{\spaceskip=\fontdimen2\font plus
\BIBentryALTinterwordstretchfactor\fontdimen3\font minus
  \fontdimen4\font\relax}
\providecommand{\BIBforeignlanguage}[2]{{%
\expandafter\ifx\csname l@#1\endcsname\relax
\typeout{** WARNING: IEEEtran.bst: No hyphenation pattern has been}%
\typeout{** loaded for the language `#1'. Using the pattern for}%
\typeout{** the default language instead.}%
\else
\language=\csname l@#1\endcsname
\fi
#2}}
\providecommand{\BIBdecl}{\relax}
\BIBdecl

\bibitem{papadopoulos2021robotic}
E.~Papadopoulos, F.~Aghili, O.~Ma, and R.~Lampariello, ``Robotic manipulation
  and capture in space: A survey,'' \emph{Frontiers in Robotics and AI}, p.
  228, 2021.

\bibitem{OOS}
A.~Flores-Abad, O.~Ma, K.~Pham, and S.~Ulrich, ``A review of space robotics
  technologies for on-orbit servicing,'' \emph{Progress in Aerospace Sciences},
  vol.~68, pp. 1--26, 2014.

\bibitem{ADR}
C.~Bonnal, J.-M. Ruault, and M.-C. Desjean, ``Active debris removal: Recent
  progress and current trends,'' \emph{Acta Astronautica}, vol.~85, pp. 51--60,
  2013.

\bibitem{OOA}
M.~Rognant, C.~Cumer, J.-M. Biannic, M.~Roa, A.~Verhaeghe, and V.~Bissonnette,
  ``Autonomous assembly of large structures in space: a technology review,'' in
  \emph{European Conf. for Aeronautics and Aerospace Sciences (EUCASS)}, 2019.

\bibitem{slotine1991applied}
J.-J.~E. Slotine, W.~Li \emph{et~al.}, \emph{Applied nonlinear control}.\hskip
  1em plus 0.5em minus 0.4em\relax Prentice hall Englewood Cliffs, NJ, 1991,
  vol. 199, no.~1.

\bibitem{parlaktuna2004adaptive}
O.~Parlaktuna and M.~Ozkan, ``Adaptive control of free-floating space
  manipulators using dynamically equivalent manipulator model,'' \emph{Robotics
  and Autonomous Systems}, vol.~46, no.~3, pp. 185--193, 2004.

\bibitem{abiko2009adaptive}
S.~Abiko and G.~Hirzinger, ``Adaptive control for a torque controlled
  free-floating space robot with kinematic and dynamic model uncertainty,'' in
  \emph{2009 IEEE/RSJ International Conference on Intelligent Robots and
  Systems}.\hskip 1em plus 0.5em minus 0.4em\relax IEEE, 2009, pp. 2359--2364.

\bibitem{wang2009passivity}
H.~Wang and Y.~Xie, ``Passivity based adaptive jacobian tracking for
  free-floating space manipulators without using spacecraft acceleration,''
  \emph{Automatica}, vol.~45, no.~6, pp. 1510--1517, 2009.

\bibitem{wang2012prediction}
------, ``Prediction error based adaptive jacobian tracking for free-floating
  space manipulators,'' \emph{IEEE Transactions on Aerospace and Electronic
  Systems}, vol.~48, no.~4, pp. 3207--3221, 2012.

\bibitem{hu2012adaptive}
Q.~Hu, L.~Xu, and A.~Zhang, ``Adaptive backstepping trajectory tracking control
  of robot manipulator,'' \emph{Journal of the Franklin Institute}, vol. 349,
  no.~3, pp. 1087--1105, 2012.

\bibitem{yu2015modeling}
X.-y. Yu and L.~Chen, ``Modeling and observer-based augmented adaptive control
  of flexible-joint free-floating space manipulators,'' \emph{Acta
  Astronautica}, vol. 108, pp. 146--155, 2015.

\bibitem{christidi2020concurrent}
O.-O. Christidi-Loumpasefski, G.~Rekleitis, and E.~Papadopoulos, ``Concurrent
  parameter identification and control for free-floating robotic systems during
  on-orbit servicing,'' in \emph{2020 IEEE International Conference on Robotics
  and Automation (ICRA)}.\hskip 1em plus 0.5em minus 0.4em\relax IEEE, 2020,
  pp. 6014--6020.

\bibitem{zhan2022extended}
B.~Zhan, M.~Jin, and J.~Liu, ``Extended-state-observer-based adaptive control
  of flexible-joint space manipulators with system uncertainties,''
  \emph{Advances in Space Research}, vol.~69, no.~8, pp. 3088--3102, 2022.

\bibitem{yao2021adaptive}
Q.~Yao, ``Adaptive trajectory tracking control of a free-flying space
  manipulator with guaranteed prescribed performance and actuator saturation,''
  \emph{Acta Astronautica}, vol. 185, pp. 283--298, 2021.

\bibitem{lee2018natural}
T.~Lee, J.~Kwon, and F.~C. Park, ``A natural adaptive control law for robot
  manipulators,'' in \emph{2018 IEEE/RSJ International Conference on
  Intelligent Robots and Systems (IROS)}.\hskip 1em plus 0.5em minus
  0.4em\relax IEEE, 2018, pp. 1--9.

\bibitem{giordano2019coordinated}
A.~M. Giordano, C.~Ott, and A.~Albu-Sch{\"a}ffer, ``Coordinated control of
  spacecraft's attitude and end-effector for space robots,'' \emph{IEEE
  Robotics and Automation Letters}, vol.~4, no.~2, pp. 2108--2115, 2019.

\bibitem{wensing2017linear}
P.~M. Wensing, S.~Kim, and J.-J.~E. Slotine, ``Linear matrix inequalities for
  physically consistent inertial parameter identification: A statistical
  perspective on the mass distribution,'' \emph{IEEE Robotics and Automation
  Letters}, vol.~3, no.~1, pp. 60--67, 2017.

\bibitem{lee2018geometric}
T.~Lee and F.~C. Park, ``A geometric algorithm for robust multibody inertial
  parameter identification,'' \emph{IEEE Robotics and Automation Letters},
  vol.~3, no.~3, pp. 2455--2462, 2018.

\bibitem{teel1996nonlinear}
A.~R. Teel, ``A nonlinear small gain theorem for the analysis of control
  systems with saturation,'' \emph{IEEE transactions on Automatic Control},
  vol.~41, no.~9, pp. 1256--1270, 1996.

\bibitem{Siciliano}
B.~Siciliano, L.~Sciavicco, L.~Villani, and G.~Oriolo, \emph{Robotics:
  Modelling, Planning and Control}, 1st~ed.\hskip 1em plus 0.5em minus
  0.4em\relax Springer Publishing Company, Incorporated, 2008.

\bibitem{armStruct}
P.~Rank, Q.~M{\"u}hlbauer, W.~Naumann, and K.~Landzettel, ``The deos automation
  and robotics payload,'' in \emph{Symp. on Advanced Space Technologies in
  Robotics and Automation, ASTRA, the Netherlands}, 2011.

\bibitem{khalil1996adaptive}
H.~K. Khalil, ``Adaptive output feedback control of nonlinear systems
  represented by input-output models,'' \emph{IEEE transactions on Automatic
  Control}, vol.~41, no.~2, pp. 177--188, 1996.

\end{thebibliography}

%%%%%%%%%%%%%%%%%%%%%%%%%%%%%%%%%%%%%%%%%%%%%%%%%%%%%%

%\section*{Acknowledgment}

%The preferred spelling of the word ``acknowledgment'' in America is without an ``e'' after the ``g''. Avoid the stilted expression ``one of us (R. B. G.) thanks $\ldots$''. Instead, try ``R. B. G. thanks$\ldots$''. Put sponsor acknowledgments in the unnumbered footnote on the first page.

\end{document}